\documentclass[sigconf]{acmart}
\usepackage{booktabs} 
\usepackage{tabularx}
\usepackage{graphicx}
\graphicspath{{figure/}}
\usepackage{subcaption}
\usepackage{amsmath,amsfonts}
\usepackage{color}
\usepackage{xspace}
\usepackage{listings}
\usepackage{breqn}
\usepackage{siunitx}
\usepackage{diagbox}
\usepackage{enumitem} 

\usepackage[font={small,sf,bf}]{caption}

\newif\iffinal

\iffinal
 \newcommand{\raj}[1]{}
 \newcommand{\zliu}[1]{}
 \newcommand{\ian}[1]{}
 \newcommand{\revision}[1]{}
\else
 \newcommand{\raj}[1]{{\textcolor{green}{ Raj: #1 }}}
 \newcommand{\zliu}[1]{{\textcolor{blue}{ Zhengchun: #1 }}}
 \newcommand{\ian}[1]{{\textcolor{red}{ Ian: #1 }}}
 \newcommand{\revision}[1]{{\textcolor{red}{#1 }}}
\fi

\newcommand{\ttfwd}{\emph{T}\/\textsuperscript{fwd}\xspace}

\newcommand{\proj}{\texttt{BFTrainer}\xspace}
\newcommand{\bfjob}{Trainer\xspace}
\newcommand{\bfjobs}{Trainers\xspace}
\newcommand{\pool}{$\mathcal{N}$\xspace}
\newcommand{\mt}{\texttt{node$\times$time}\xspace}
\newcommand{\nodehour}{node-hours\xspace}
\newcommand{\nodesecond}{node-seconds\xspace}
\newcommand{\tfwd}{forward-looking time\xspace}
\newcommand{\pjmax}{\emph{Pj}\textsubscript{max}}

\newcommand{\summitkillable}{\texttt{killable}\xspace}
\newcommand{\dps}{\displaystyle}
\newcommand{\st}{\mbox{subject to }}
\newcommand{\maxi}{\mathop{\mbox{maximize}}}

\theoremstyle{definition} 
\newtheorem{obs}{Observation}
\usepackage{tcolorbox,tikz}
\tcbuselibrary{skins}
\tcbuselibrary{breakable}
\newtcolorbox{obsbox}[3][]
{
  breakable, 
  enhanced,
  colback  = #2!2,
  colframe= #2!8,
  boxsep=-0.5mm,
  borderline west={1.5mm}{0.05mm}{#3!15}, 
  #1,
}

\setcopyright{acmcopyright}
\copyrightyear{2021}
\acmYear{2021}
\acmDOI{}

\acmConference[Preprint]{XXX YYY}
\acmBooktitle{Under peer review}
\acmPrice{15.00}
\acmISBN{978-1-4503-XXXX-X/18/06}

\setcopyright{none}
\renewcommand\footnotetextcopyrightpermission[1]{}

\begin{document}

\title{BFTrainer: Low-Cost Training of Neural Networks on Unfillable Supercomputer Nodes} 
\author{Zhengchun Liu}
\affiliation{%
  \institution{Argonne National Laboratory}
}
\email{zhengchun.liu@anl.gov}
\orcid{0000-0002-6647-4423}

\author{Rajkumar Kettimuthu}
\affiliation{\institution{Argonne National Laboratory}}
\email{kettimut@anl.gov}

\author{Michael E. Papka}
\affiliation{\institution{Argonne National Laboratory}}
\affiliation{\institution{Northern Illinois University}}
\email{papka@anl.gov}

\author{Ian Foster}
\affiliation{\institution{Argonne National Laboratory}}
\affiliation{\institution{The University of Chicago}}
\email{foster@anl.gov}

\begin{abstract}
Supercomputer FCFS-based scheduling policies result in many transient idle nodes, a phenomenon that is only partially alleviated by backfill scheduling methods that promote small jobs to run before large jobs. 
Here we describe how to realize a novel use for these otherwise wasted resources, namely, deep neural network (DNN) training.
This important workload is easily organized as many small fragments that can be configured dynamically to fit essentially any node$\times$time hole in a supercomputer's schedule.
We describe how the task of rescaling suitable DNN training tasks to fit dynamically changing holes can be formulated as a deterministic mixed integer linear programming (MILP)-based resource allocation algorithm, and show that this MILP problem can be solved efficiently at run time.
We show further how this MILP problem can be adapted to optimize for administrator- or user-defined metrics.
We validate our method with supercomputer scheduler logs and different DNN training scenarios, and demonstrate efficiencies of up to 93\% compared with running the same training tasks on dedicated nodes.
Our method thus enables substantial supercomputer resources to be allocated to DNN training with no impact on other applications.  
\end{abstract}


\maketitle
\pagestyle{plain}

\section{Introduction}\label{sec:introduction}
Supercomputers typically service requests for computational resources 
with policies that adhere roughly to first come first serve (FCFS) semantics. 
This approach inevitably results in idle nodes as larger tasks block subsequent tasks from executing. 
Backfilling~\cite{srinivasan2002characterization}, a strategy by which later tasks are promoted to run sooner if doing so does not delay other tasks, can improve efficiency, but substantial idle resources inevitably remain.
For example, if tasks T1, T2, and T3 request 20\%, 90\%, and 50\% of all nodes, respectively, FCFS will run these tasks sequentially, in that order, with the result that 80\% and 10\% of the machine is idle as T1 and T2 execute, respectively.  
If T3 requires less time than T1, then backfilling can promote it to run concurrently with T1, but substantial idle resources remain.

Such unfillable nodes are 
common on leadership-class supercomputers that operate under policies that prioritize large applications that cannot run on small clusters or would take so long as to be impractical: what is known as capability computing~\cite{allcock2017experience,ics20-alcf-logs,patel2020job}.
We refer to a node that the main scheduler does not use for (regular or backfilled) jobs as an \emph{idle node}, and denote the set of all idle nodes at a particular time as \pool{}. 

We describe here how \pool{} can be used effectively for 
DNN training, an 
activity that consumes growing numbers
of compute resources in the cloud~\cite{qiao2020pollux} and at supercomputer centers~\cite{you2018imagenet}.
DNN training is malleable due to the data parallelism mechanism used to process a batch of data for gradient calculation, and can easily be rescaled as 
one needs to checkpoint only the model weights and (in the case of stateful optimizers) optimizer state. 
Indeed,  deep learning frameworks such as AdaptDL~\cite{qiao2020pollux}, TorchElastic of PyTorch~\cite{paszke2019pytorch}, and Elastic Horovod~\cite{sergeev2018horovod} enable scaling up and down the number of workers dynamically at runtime with slight cost, without requiring a restart or resuming from checkpoints saved to durable storage.

Our proposed \proj{} leverages this malleability of DNN training to make optimal use of non-backfilled idle supercomputer nodes. 
Basically, \proj{} collects idle nodes into a resource pool, \pool{}, from which it reallocates them for DNN training jobs 
(referred to as \bfjob{}s in the rest of the paper). 
The core idea is that because nodes in \pool{} can come and leave without any commitment,
the use of \proj{} does not affect jobs submitted to the main scheduler.
That same characteristic makes it difficult for conventional HPC applications that are not malleable, and are expensive to migrate or checkpoint and restart, to use \pool{} effectively. 

\proj{} may be used in two  ways.
In the first, a single user runs multiple \bfjob{}s: for example, for hyperparameter optimization (HPO) or neural architecture search (NAS). 
Here each user has an isolated \proj{} instance and specifies a metric (e.g., aggregated throughput) for \proj{} to allocate nodes optimally for their \bfjob{}s. 
To support multiple users or if a single user 
cannot consume all idle nodes, we can divide the entire supercomputer into several small clusters logically and  have one \proj{} for each small cluster to manage idle nodes for that cluster.
In the second scenario, 
all users submit to a single \proj{} instance;
in this case, an
administrator needs to propose a priority score (e.g., efficiency) for \proj{} to plan resource allocation optimally. 

Specific benefits of \proj{} include the following:
\begin{itemize}
\item \textbf{Zero cost to jobs in the main scheduler:} All \bfjob{}s running on idle nodes are fully preemptable at all times; thus, \proj{} has no impact on other jobs. 
\item \textbf{Optimal elastic scheduling}: A mathematical optimization model is used to 
identify node allocations for \bfjob{}s that are optimal given available information. 
\item \textbf{Customizable objective}: Administrators or users can assign a metric, for example, throughput, scaling efficiency, or priority score, and \proj{} will allocate resources to different \bfjob{}s to optimize the metric.
\end{itemize}

The use of \proj{} can deliver benefits to both supercomputer centers and supercomputer users.
For a supercomputer center, \proj{} can increase the amount of computing performed;
for users, assuming that their supercomputer center incentivizes the use of \proj{} by reduced charges, it can reduce their costs if they are prepared to accept a less certain scheduling approach (e.g., AWS spot instances provide a similar tradeoff). Note that depending on relative demand, \proj{} may run slower or faster than the main queue.


We next use Summit supercomputer logs to obtain quantitative data on the frequency of unfillable nodes (\S\ref{sec:real-data-analysis}); 
define an MILP model for allocating nodes in \pool{} to \bfjob{}s (\S\ref{sec:mip-mdl-form});
introduce our experimental setup and evaluation metrics (\S\ref{sec:exp-setup});
evaluate our resource allocation algorithm by replaying real traces with real DNN models (\S\ref{sec:exp-ana});
review related work (\S\ref{sec:related-work}); and conclude (\S\ref{sec:conclusion}).


\section{Situation in Practice}\label{sec:real-data-analysis}
Our first concern is the characteristics of idle nodes of modern supercomputers and the limitations of current implementations. 
Specifically, we want to answer the following questions before designing the resource allocation scheme:
What is the quantity of idle nodes on average? Is it large enough to motivate having the \proj{}?
What is the typical lifetime of an idle node?
How frequently does \pool{} change (nodes join and leave)?

\subsection{Idle-node characterization}\label{sec:spans-characterization}
We studied idle nodes on Summit, a \num{4608}-node supercomputer at Oak Ridge National Laboratory
that uses IBM Spectrum Load Sharing Facility (LSF) as its batch scheduler.
To identify idle nodes, we ran, every 10 seconds, the LSF
\texttt{jobstat} command to identify jobs that are currently running or scheduled to run, and \texttt{bslots} to inspect backfill windows and identify currently idle nodes.
We remove from the resulting list of idle nodes those listed as idle in the period between when one job releases a node and the next job starts to use that node (these usually appear only in one \texttt{bslots} record). 
We then generate events by removing records for which there is no change in \pool{} from one record to the next.

We monitored the state of each node continuously for two months; see \autoref{tbl:char-idle-nodes}.
Here we study the characteristics of the idle state of each node for the first two weeks that we will use for experiments in~\S\ref{sec:exp-setup} and \S\ref{sec:exp-ana}. 
Integrating the number of nodes over time, we determined that Summit nodes are idle for a total of \num{137639} \nodehour{} over the two weeks, equivalent to 407 nodes or 8.6\% of the whole machine being idle for two weeks.
For ease of reference, we refer to each change in the composition of \pool{}, whether by nodes joining and/or leaving, as an \textit{event}. 
If multiple nodes join or leave \pool{} at the same time, we treat this as one event. 
We use \textit{fragment} to represent a period during which a particular node is idle; thus, one physical node may lead to many fragments in the logs at different times. 
We identified \num{22883} events ($\sim$68 per hour), during two arbitrary weeks:
\num{14049} in which at least one node joined \pool{} ($\sim$42 per hour) and
\num{10573} in which at least one node left \pool{} ($\sim$31 per hour).

\begin{figure}[htb]
\centering
\vspace{-.2cm}
\includegraphics[width=0.8\columnwidth]{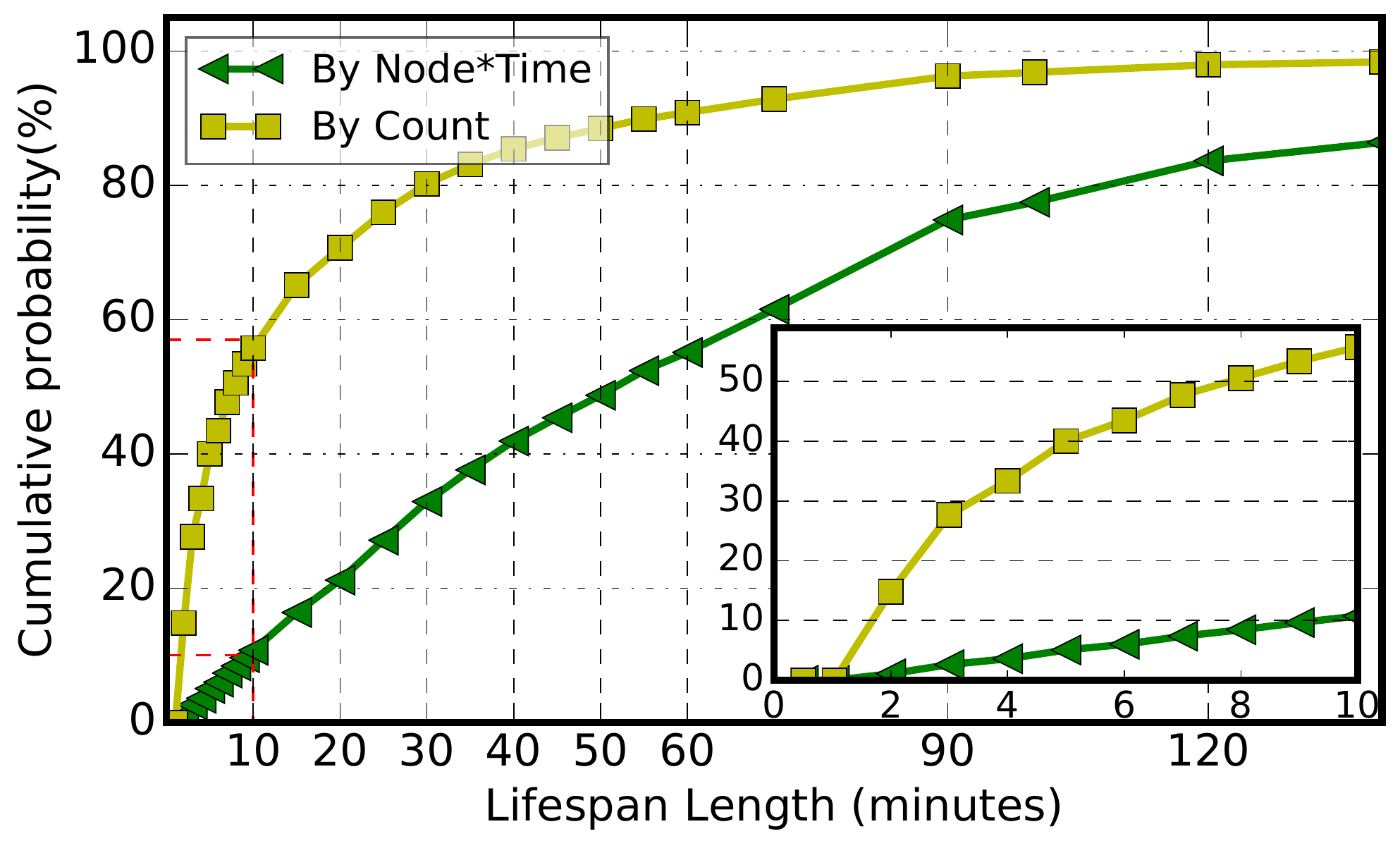}
\caption{Cumulative distribution of fragment length.}
\vspace{-.2cm}
\label{fig:spans-cdf}
\end{figure}

Besides the frequency of the change to \pool{}, the length of fragments, which indicates the granularity of resources, is also important for the design of the resource allocation algorithm. 
\autoref{fig:spans-cdf} shows a cumulative distribution of the fragment length.
We see that most idle node times are short: for example, about 58\% of the fragments are idle  less than 10 minutes. 
Although those short fragments  account for only a small portion of all idle resources in \mt{} (e.g., these 58\% short node-spans contribute to only about 10\% of total idle \mt{}), there is a static cost in scaling up applications in order to make use of each node span. 
Moreover, because of the stochastic pattern of job submission and the uncertainties of wall-time estimation, we cannot know the size of fragment at the time when a node joins \pool{}. 
Thus, these short fragments may lead us to pay the rescaling cost to scale up \bfjobs{} when nodes join \pool{} and then soon be forced to scale down (with a cost) by preemption and end up with no progress on \bfjobs{}.

\autoref{tbl:char-idle-nodes} summarizes idle nodes on three leadership-class supercomputers based on their scheduler logs. 
The number of idle nodes changes less frequently on \texttt{Theta} and \texttt{Mira} (both on full 2019 logs) than on \texttt{Summit} (2021 Feb. and Mar.), because the former machines constrain the minimum number of nodes that can be requested (128 and 512 for Theta and Mira, respectively) in order to encourage large-scale jobs that cannot be run elsewhere~\cite{ics20-alcf-logs,allcock2017experience}, whereas Summit allows requests of a single node.
Constraining the minimum number of nodes leads to fewer changes, but more idle resources. 

\begin{table}[htb]
\centering
\caption{Characteristics of resources that cannot be backfilled. \texttt{INC/h} and \texttt{DEC/h} indicate the average number of times per hour that idle nodes increased and decreased, respectively. \texttt{eq-Nodes} denotes the number of nodes that would need to be available continuously to deliver resources equal to the idle resources in \mt{}.}
\begin{tabular}{c|c|c|c|r|c}
\noalign{\hrule height 2pt}
\textbf{System} & \textbf{Duration} & \textbf{INC/h} & \textbf{DEC/h} & \multicolumn{1}{c|}{\textbf{Ratio}} & \textbf{eq-Nodes}\\\hline
Summit & 2 months & 41.7 & 28.6 & 11.1\% & 524 \\\hline
Theta & 1 year & 6.3 & 6.2 & 12.5\% & 547\\\hline
Mira & 1 year & 2.8 & 2.4 & 10.3\% & 5071\\
\noalign{\hrule height 2pt}
\end{tabular}
\label{tbl:char-idle-nodes}
\end{table}

These availability characteristics make it challenging to use such fragmented resources efficiently because there is always a cost to rescale/restart an application. 
For example, if scaling up a job takes 20 seconds and a job is running on 10 nodes, then adding one idle node to that job will result in 10 nodes waiting 20 seconds as the new node is initialized and integrated into the job: a cost of 200 \nodesecond{}. 
Thus, an efficient strategy is needed to decide, whenever \pool{} changes, whether and how to rescale each \bfjob{}.


\begin{obsbox}[boxsep=1pt,left=8pt,right=2pt,top=0pt,bottom=0pt]{green}{green}
\obs{On average, about 10\% of nodes cannot be backfilled, a number that is larger
on supercomputers that impose large minimum job sizes to encourage capability computing. 
Most idle fragments are short and contribute little \mt{}; for example, 58\% 
are smaller than 10 minutes and contribute only 10\% of total \mt{} on Summit.
Short fragments are challenging to use because there is a fixed cost to rescale a running job and we cannot know fragment duration. 
}
\end{obsbox}

\subsection{Existing implementations}
The \summitkillable{} queue of Summit is a preemptable queue that allows jobs within a certain size constraint to request wall time of up to 24 hours. 
A job submitted to this queue become preemptable once it reaches the guaranteed runtime limit, which depends on job size. 
\summitkillable{} is beneficial to applications, including deep learning training, that can easily restart from a checkpoint. 
Assume a job $\widehat{J}$ submitted to the \summitkillable{} queue with size of $\widehat{N}$ asking for a guaranteed runtime of $\widehat{T}$ started to run at time $t$=0.
The key differences between \summitkillable{} and \proj{} are as follows: 
\begin{enumerate}
\item $\widehat{J}$ can use only exactly $\widehat{N}$ nodes, that is, any additional idle nodes from $t$=0 to $\widehat{T}$ cannot be used even if $\widehat{J}$ is malleable.
\item If any nodes that were allocated to $\widehat{J}$ are not available after $\widehat{T}$, the job will be killed instead of being rescaled or migrated.
\item \proj{} collects all idle nodes into a \pool{} and allocates them to multiple \bfjobs{} so that the goal (e.g., utilization efficiency, aggregated throughput) is optimized.
\end{enumerate}

In summary, the scheduling strategies used on leadership-class systems to support capability computing unavoidably result in considerable idle but fragmented resources. 
\proj{} is designed to improve the utilization of such systems without impacting the resources delivered to other jobs (whether currently in the queue or to be submitted in the future) by the main scheduler.
Other utilization-improving solutions such as dynamic priorities between users and groups~\cite{abraham2015priority,steffen2019better} are complementary.

\section{MILP Model}\label{sec:mip-mdl-form}
As noted in \S\ref{sec:spans-characterization}, nodes may join \pool{} at any time, without any commitment, and may leave \pool{} at any time, without warning. 
From the point of view of a \bfjob{}, a preemption cost is unavoidable when one or more of the nodes that it is using leave, and there is always a cost when scaling a \bfjob{} up or down. 
Thus, we face the challenge of determining an allocation of nodes to \bfjob{}s at each event or when a \bfjob{} completes, in order to maximize the return (e.g., aggregated throughput). 

\begin{figure}[htb]
\vspace{-.3cm}
\center
\includegraphics[width=0.8\columnwidth]{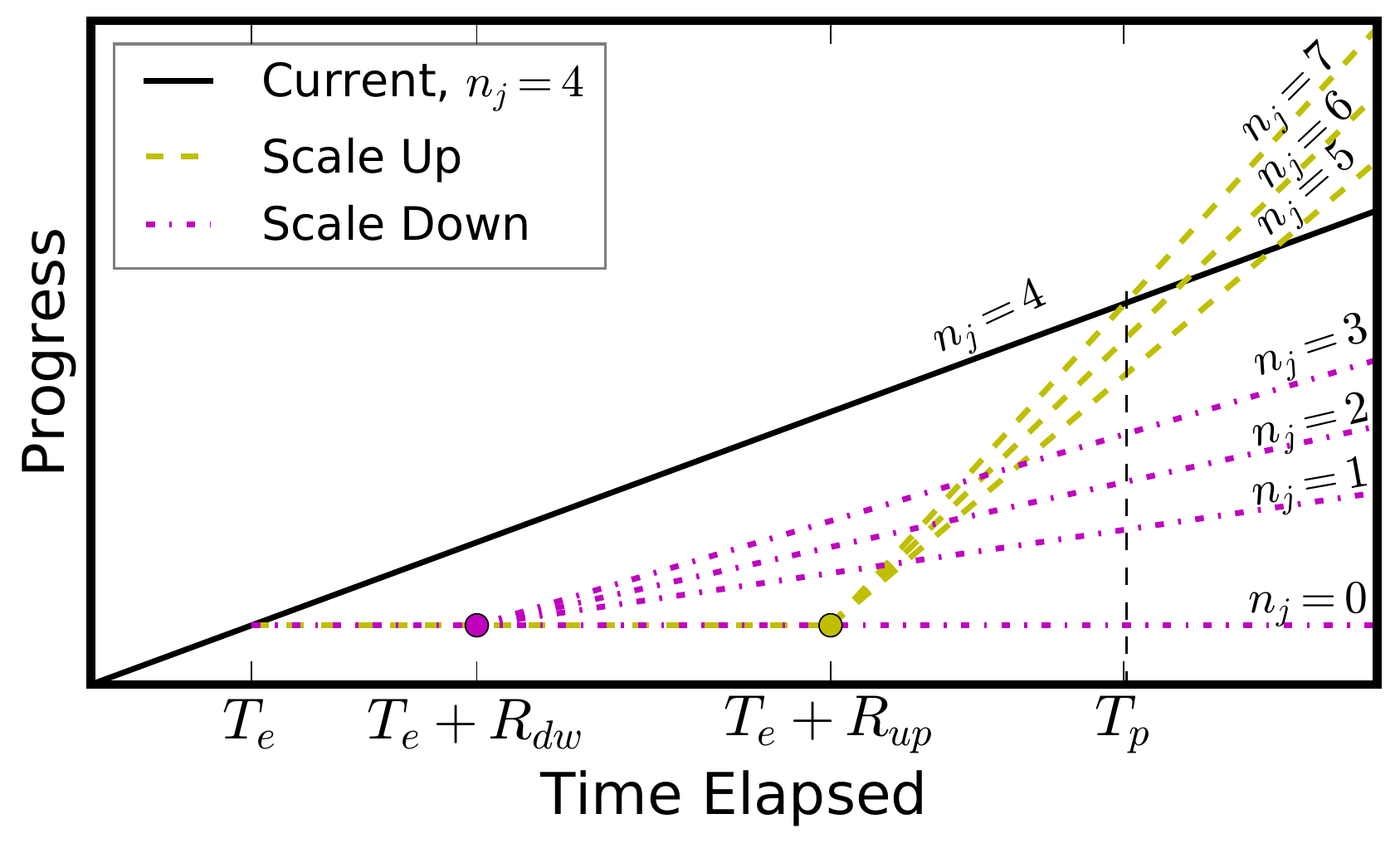}
\caption{The $|$\emph{S}\textsubscript{\emph{j}}$|$ decision choices that arise when rescaling a \bfjob{}, shown here for the case $|$\emph{S}\textsubscript{\emph{j}}$|$=7.} 
\label{fig:decision}
\vspace{-.4cm}
\end{figure}

\begin{figure}[htb]
\center
\vspace{-.3cm}
\includegraphics[width=0.8\columnwidth, trim=0cm 0 0.2cm .2cm, clip]{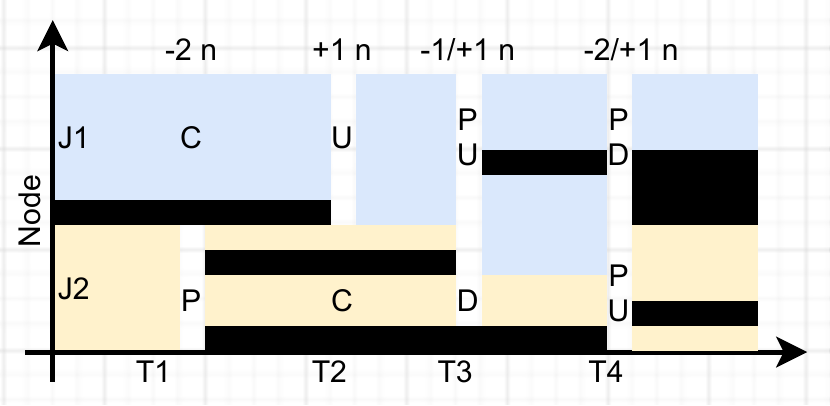}
\caption{Event-driven resource reallocation and rescaling. \texttt{C}:~Continue; \texttt{P}: Preemption; \texttt{U}: Upscaling; \texttt{D}: Downscaling. \texttt{Black fill}: main scheduler occupancy; \texttt{White gap}: rescaling/preemption cost.} 
\label{fig:demo}
\vspace{-.4cm}
\end{figure}

\autoref{fig:decision} demonstrates the rescaling decision choices that arise for a \bfjob{} when nodes leave \pool{} at time $T_e$, forcing some \bfjob{}s to scale down and thus creating a potentially suboptimal use of nodes.
There are $J$ \bfjob{}s in \proj{}, and for each we may do the following: 
(i) not change the number of nodes: incurs no rescaling cost;
(ii) scale up (other \bfjob{}s must scale down to release nodes): incurs cost $R_{up}$; or
(iii) scale down (will scale up others to use the released nodes): incurs cost $R_{dw}$.
The task is the same if nodes join \pool{} at time $T_e$.
\autoref{fig:demo} illustrates decisions that \proj{} needs to make in different scenarios (two \bfjob{}s try to fill fragments out of 11 nodes).
Since each \bfjob{} can run on a number of nodes $n_j \in S_j = \{0, N_{j}^{min}..N_{j}^{max}\}$, where $n_j=0$ corresponds to setting the \bfjob{} to a ``waiting'' state, there are a total of $|S_j|$ decision choices for each \bfjob{}.
Although it is straightforward to quantify the gain and loss associated with each choice (i.e., the slope that denotes the rate of progress increases or decreases), the search space will be $|S_1| \times |S_2| \ldots |S_J|$ (i.e., $\prod_{j=1}^{J}|S_j|$) to find the optimal decision to maximize the total throughput of $J$ \bfjob{}s.
This exponentially increasing search space leads us to seek a deterministic global optimization solution from mathematical programming solutions.

Linear programming (LP)
is a technique for  optimizing a linear objective function, subject to linear equality and linear inequality constraints. 
Mixed-integer linear programming (MILP), a subset of LP, involves problems in which some variables are constrained to be integers, while others may be nonintegers.
A LP problem comprises three elements: \textit{decision variables}, quantities that need to be determined in order to solve the optimization problem; an \textit{objective function}, which provides a criterion to be maximized or minimized; and \textit{constraints}, which describe the limitations that restrict choices of decision variables.
All constraints and the objective function must be linear and continuous in order to be solvable.  

Here, we optimize resource allocation by solving a MILP problem where the decision variable is the allocation of each node and the objective function is the metric (e.g., total throughput) to optimize. 
In practice, we solve a MILP whenever there is a change to \pool{}, a \bfjob{} completes, or a new \bfjob{} is ready to run.
We detail the three elements (decision variables, constraints, and objective function) of our MILP problem in \S\ref{sec2:variable}, \S\ref{sec2:constraint}, and \S\ref{sec2:objective} respectively.

\subsection{Definition of Sets and Indices}
We represent the resources managed by \proj{} by three sets:
\begin{description}
\item[Nodes$, {\mathcal N},$] is the set of idle nodes. 
\item[Jobs$, {\mathcal J},$] is the set of \bfjob{}s  running in \proj{}. 
\item[Event$, {\mathcal E},$] is the events ($e_1, e_2, \ldots, e_i, \ldots$) in which nodes join and/or leave \pool{} at time $T_{e_i}$.
\end{description}

\noindent
We represent a \bfjob{} $j,$ $j \in \mathcal{J}$ by the following variables.
\begin{description}
\item[$N_{j}^{min}$ and $N_{j}^{max}$] are the minimum and maximum number of nodes on which \bfjob{} $j$ can run.
\item[Job Scale$, N_{j}$] is the number of nodes on which \bfjob{} $j$ is currently running.
\item[Throughput$, A_{thr}$,] is a measure of application throughput: for deep learning model training, we use samples per second. 
\item[Scale Up$, R_j^{up} $] is the time in seconds to scale up \bfjob{} $j$; i.e., the time that $N_{j}$ nodes sit idle as the new node(s) are prepared (clone current model and initialize data pipeline).
\item[Scale Down$, R_j^{dw}$,] the time in seconds to scale down \bfjob{} $j$, is usually smaller than $R_j^{up}$.
\item[Objective Metric$, O_j\left(N_j \right)$,] a function of \bfjob{} scale ($N_j$), denotes the gain (toward the objective) obtained per second if \bfjob{} $j$ is run on $N_j$ nodes. 
$O_j\left(N_j \right)$ is agnostic to weak or strong scaling. A \bfjob{} can either fix global batch size by adjusting batch size per node or fix the batch per node with dynamic global batch. In the latter case, we assume that the user employs strategies, e.g.,  Adasum~\cite{maleki2021scaling}, for dynamic global batches.
\end{description}

Values must be supplied for $N_{j}^{min}$, $N_{j}^{max}$, $R_j^{up}$, and $R_j^{dw}$. 
The function $O_j\left(N_j \right)$ (e.g., throughput) can be determined by \proj{} by 
experiment; alternatively, it can be provided by the user.

\begin{description}
\item[BigM$,M,$] is a sufficiently large constant ($>|\mathcal{N}|$) used for modeling purposes. 
\item[Nodes to \bfjob{} map$,c_{jn},$] is a 2D binary matrix, indexed by $j$ and $n$;
each element is 1 iff node $n$ ($n \in \mathcal{N}$) is allocated for \bfjob{} $j$ ($j \in \mathcal{J}$). 
\end{description}

\subsection{Decision Variables}\label{sec2:variable}
The decision variable $x_{jn}$ is binary, with value 1 iff node $n$ is allocated for \bfjob{} $j$.
Basically, solving the MILP problem transfers $c_{jn}$ to $x_{jn}$ to maximize the objective (gain integral) for each event $e\in\mathcal{E}$;
the number of nodes allocated to \bfjob{} $j$ changes from
\begin{equation}
C_j = \sum_{n \in {\mathcal N}}c_{jn} \quad \forall j \in \mathcal{J}
\end{equation}
to
\begin{equation}
N_j = \sum_{n \in {\mathcal N}}x_{jn} \quad \forall j \in \mathcal{J}
\end{equation}
after the event.

\subsection{Constraints}\label{sec2:constraint}
\subsubsection{Job size constraints}
The number of nodes allocated to \bfjob{} $N_j$  must satisfy $N_j = 0$ (i.e., the \bfjob{} sits idle) or $N_j^{min} \le N_j \le N_j^{max}$.
Logically, it is equivalent to
\begin{equation}
\begin{aligned}
N_j^{min} \le N_j  \quad || \quad  N_j = 0\\ 
N_j \le N_j^{max}  \quad || \quad N_j = 0 \\
\forall j \in {\mathcal J}.
\label{eq:math-cnt-jsz}
\end{aligned}
\end{equation}

We then reformulate the nonlinear constraints described as \autoref{eq:math-cnt-jsz} into \autoref{eq:cnt-jsz} to satisfy the requirements of MILP:
\begin{equation}
\begin{aligned}
N_j \ge N_j^{min} - My_j^l, \\
N_j \le M(i-y_j^l), \\
N_j^{max} \ge N_j - My_j^u, \\
N_j \le M(1 - y_j^u), \\
y_j^l, y_j^u \in \{0, 1\}, \forall j \in {\mathcal J},
\label{eq:cnt-jsz}
\end{aligned}
\end{equation}
where $y_j^l$ and $y_j^u$ are auxiliary (binary) decision variables.

\subsubsection{Node allocation constraints}
As each node can  be allocated only to at most one \bfjob{}, we have constraints 
\begin{equation}
\sum_{j \in {\mathcal J}}x_{j}^{n} \le 1 \quad \forall n \in {\mathcal N}.
\label{eq:nodes-constraint}
\end{equation}

\subsubsection{Job migration constraints}
Since the migration of a \bfjob{} from one node to another will always introduce a cost and provides no benefit in this study, we include constraints to disallow it:
\begin{equation}
\sum_{n \in \mathcal{N}}x_{jn} \oplus c_{jn} = \left | \sum_{n \in \mathcal{N}}x_{jn} - \sum_{n \in \mathcal{N}}c_{jn} \right | \quad \forall j \in \mathcal{J},
\label{eq:disable-migr-concpt}
\end{equation}
where $\oplus$ denotes the \texttt{XOR} (exclusive or) operation. 
\autoref{eq:disable-migr-concpt} is equivalent to 
\begin{equation}
\begin{aligned}
\sum_{n \in \mathcal{N}}x_{jn} - \sum_{n \in \mathcal{N}}c_{jn} \ge \sum_{n \in \mathcal{N}}x_{jn} \oplus c_{jn} \\
\text{OR} \\
\sum_{n \in \mathcal{N}}x_{jn} - \sum_{n \in \mathcal{N}}c_{jn} \le -\left(\sum_{n \in \mathcal{N}}x_{jn} \oplus c_{jn}\right)\\
\forall j \in \mathcal{J},
\end{aligned}
\end{equation}
which can further be formulated as linear constraints 
\begin{equation}
\begin{aligned}
\sum_{n \in \mathcal{N}}x_{jn} - \sum_{n \in \mathcal{N}}c_{jn} \ge \sum_{n \in \mathcal{N}}x_{jn} \oplus c_{jn} - Mz_j, \\
\sum_{n \in \mathcal{N}}x_{jn} - \sum_{n \in \mathcal{N}}c_{jn} \le -\sum_{n \in \mathcal{N}}x_{jn} \oplus c_{jn} + M(1-z_j)\\
z_j \in \{0, 1\}, \forall j \in {\mathcal J}
\label{eq:migration-constraint-ccpt}
\end{aligned}
\end{equation}
by introducing an auxiliary (binary) variable $z_{j} \in \{0, 1\}$.

To reformulate the \texttt{XOR} operation of variable $x_{jn}$ and constant $c_{jn}$, we define auxiliary variable $u_{jn} \in \{0, 1\}, \forall j \in {\mathcal J}, \forall n \in {\mathcal N}$ to represent $x_{jn} \oplus c_{jn}$, and introduce these constraints:
\begin{equation}
\begin{aligned}
u_{jn} \le x_{jn} + c_{jn} \\
u_{jn} \ge x_{jn} - c_{jn} \\
u_{jn} \ge c_{jn} - x_{jn} \\
u_{jn} \le 2 - x_{jn} - c_{jn} \\
u_{jn} \in \{0, 1\}, \forall j \in {\mathcal J}, \forall n \in {\mathcal N}.
\label{eq:xor}
\end{aligned}
\end{equation}

Thus, we get the final constraints to disallow \bfjob{} migration by substituting $u_{jn}$ into \autoref{eq:migration-constraint-ccpt}, leading to
\begin{equation}
\begin{aligned}
\sum_{n \in \mathcal{N}}x_{jn} - \sum_{n \in \mathcal{N}}c_{jn} \ge \sum_{n \in \mathcal{N}}u_{jn} - Mz_j, \\
\sum_{n \in \mathcal{N}}x_{jn} - \sum_{n \in \mathcal{N}}c_{jn} \le -\left(\sum_{n \in \mathcal{N}}u_{jn} \right) + M(1-z_j)\\
\forall j \in {\mathcal J}.
\label{eq:migration-constraint}
\end{aligned}
\end{equation}

\subsection{Objective Function}\label{sec2:objective}
The objective function has two parts: the calculation of the objective function value for the new resource allocation map and the cost to rescale (up or down) \bfjob{}s.
The new objective is calculated based on an expected steady time, \ttfwd{}, referred to as \tfwd{}, during which we assume no nodes will leave \pool{} (i.e., no \bfjob{} will be forced to scale down).
The cost to rescale \bfjob{}s is a one-time cost regardless of \ttfwd{}.

\subsubsection{Objective metric approximation}
\begin{figure}[htb]
\vspace{-.3cm}
\center
\includegraphics[width=0.8\columnwidth]{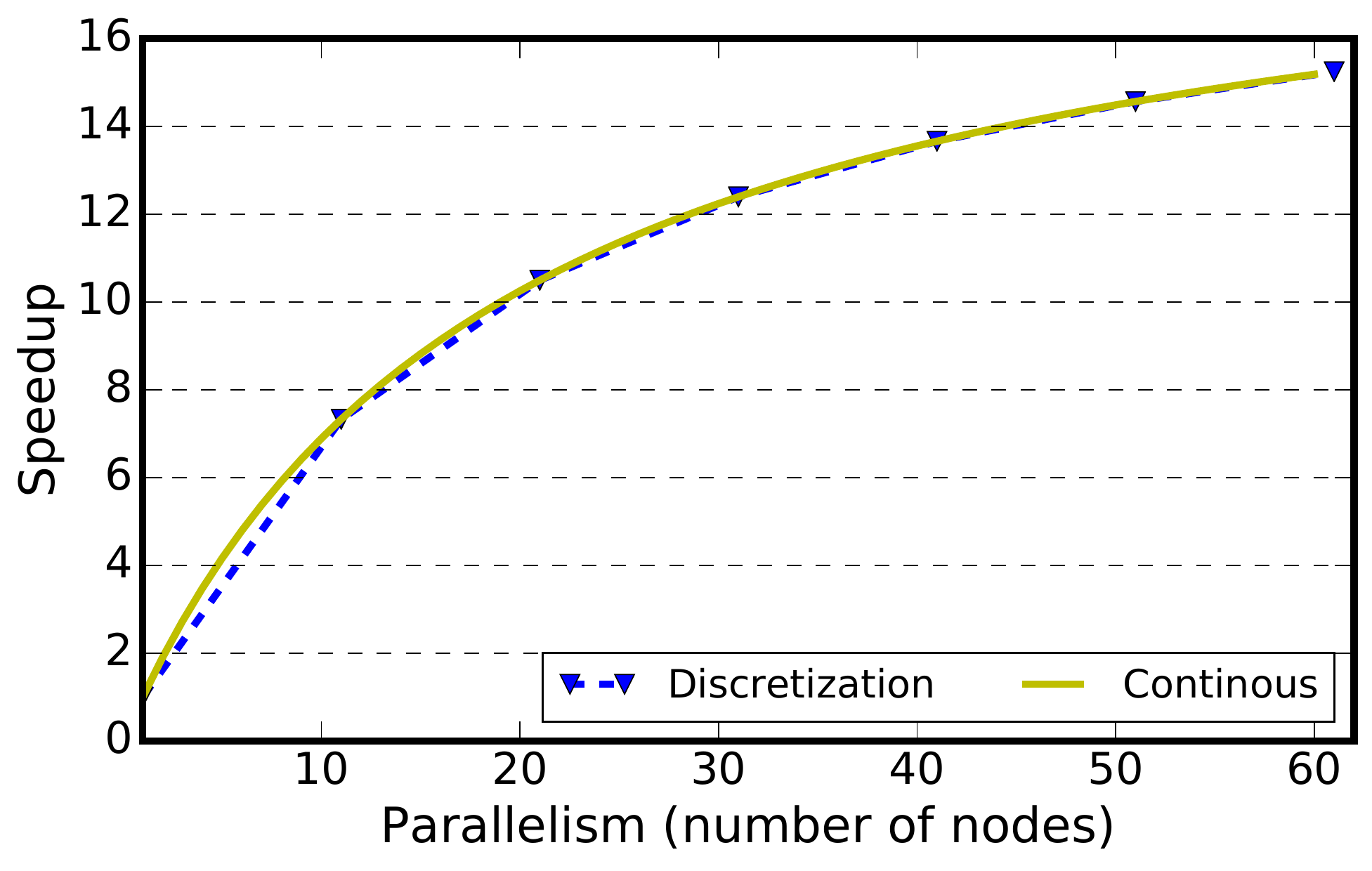}
\caption{Example of approximating parallel application scalability.} 
\label{fig:Amdahl}
\vspace{-.4cm}
\end{figure}

A \bfjob{}'s objective metric, such as scalability, is usually a nonlinear function.
Consider a nonlinear scalability function $O_j(N_j)$ that models the throughput of \bfjob{} $j$ when run with  different numbers of nodes ($N_j^{min} \le N_j \le N_j^{max}$).
The graph in \autoref{fig:Amdahl} depicts the growing of $O_j(N_j)$ for $N_j \in [1, 60]$. 
Triangles indicate selected discretization points for $N_j$. 
We observe that in this case the approximation (straight lines connecting the triangles) remains close to the real curve using only just a few discretization points. 
Additional discretization points can be included, not necessarily evenly distributed, for improved precision.

To enable the MILP model to work with these functions, we use 
special ordered sets (SOSs) to obtain linear approximations. 
SOSs are ordered sets of variables in which only one or two contiguous variables can assume nonzero values~\cite{beale1970special}. 
They provide powerful means of modeling nonconvex functions~\cite{beale1976global} and can improve the performance of the branch-and-bound algorithm.
Type~2 SOS (S2), a set in which up to two consecutive variables may assume nonzero values, 
are especially useful for modeling piecewise linear approximations of nonlinear functions.
Here we use S2 to model or approximate the scalability function of $O_j(N_j)$.

To approximate $O_j(N_j)$, for each \bfjob{} $j \in \mathcal{J}$, we discretize its scalability into $D_j$-1 sections, in other words, for consecutive points $N_j \in \{N_j^1 <, \ldots, < N_j^{D_j}  \}$, which correspond to $O_j=O_j(N_i) \in \{s_j^1, \ldots, s_j^D \}$. 
We define auxiliary continuous variables $0.0 \le w_j^1, \ldots, w_j^D \le 1.0 \quad \forall j \in \mathcal{J}$ (i.e., this optimization becomes MILP because of the variable $w$).
Then, for any given $n_x \in [N_j^{min}, N_j^{max}]$, the approximation of $O(n_x)$ could be computed with the following constraints over $n, s$, and $w_j^1,\ldots, w_j^D$:
\begin{equation}
\begin{aligned}
\sum_{i \in \mathcal{D}}w_j^i=1  \\
\sum_{i \in \mathcal{D}}w_j^iN_j^i = n_x\\
0.0 \le w_i^i \le 1.0, \forall i \in {\mathcal D}_j, \forall j \in \mathcal{J},
\label{eq:scalability-sos2}
\end{aligned}
\end{equation}
provided that $\{(w_1, n_1), \ldots, (w_D, n_D)\}$ is included as an S2 in the solver, where ${\mathcal D}_j = \{1, \ldots, D_j\}$.
The approximated $O_j(n)$ is then:
\begin{equation}
O_j(n) = \sum_{i \in {\mathcal D}_j}w_j^is_j^i.
\end{equation}

\subsubsection{Rescaling cost}
Scaling up is usually more expensive than scaling down because of the significant initialization (e.g., data pipeline) costs incurred on the new nodes.
Basically the rescaling cost, $R_j$ in seconds, for \bfjob{} $j$ will be
\begin{equation}
\begin{aligned}
R_j=\left\{\begin{matrix}
R_j^{dw} \quad N_j < C_j\\
0 \quad \quad N_j = C_j\\
R_j^{up} \quad N_j > C_j 
\end{matrix}\right.,
\label{eq:cost-step-funx}
\end{aligned}
\end{equation}
where $R_j^{dw}$ and $R_j^{up}$ are \bfjob{}-dependent scale-down and scale-up costs, respectively; $N_j = \sum_{n \in \mathcal{N}}x_{jn}$ is the number of nodes to be allocated to \bfjob{} $j$; and $C_j = \sum_{n \in \mathcal{N}}c_{jn}$ is the number of nodes on which \bfjob{} $j$ is currently running. 
We introduce two auxiliary variables, $z^u_j, z^d_j \in \{0, 1\}$, for each \bfjob{} to represent scale-up and scale-down costs, respectively, and reformulate \autoref{eq:cost-step-funx} as
\begin{equation}
R_j=\sum_{j \in \mathcal{J}} \left( R_j^{up}z^u_j + R_j^{dw}z^d_j  \right),
\label{eq:rescale-cost}
\end{equation}
which is constrained to 
\begin{equation}
\begin{aligned}
N_j \le C_j + (M - C_j)z^u_j\\
N_j \ge (C_j + 1)z^u_j\\
N_j \le (C_j-1)+\left(M-(C_j-1)\right)(1-z^d_j)\\
N_j \ge C_j(1-z^d_j)\\
\forall j \in \mathcal{J}
\end{aligned}
\end{equation}
to satisfy linear programming.
If the objective function is an integration of time (e.g., total outcome in \ttfwd{}), we will need to integrate the \bfjob{} throughput over the cost on time (\autoref{eq:rescale-cost}) to compute the rescaling cost in the objective function.

\subsubsection{Objective function}
The final objective function will be
\begin{equation}
\begin{aligned}
\sum_{j \in \mathcal{J}}T^{\text{fwd}}f_j(N_j) - \sum_{j \in \mathcal{J}}f_j(C_j)R_j ,
\label{eq:objf}
\end{aligned}
\end{equation}
where \ttfwd{} is the forward-looking time during which we assume no nodes leave \pool{}. 
If \ttfwd{} is too optimistic, the decision will not be optimal because 
\bfjob{}s will be forced to scale down due to preemption.
If \ttfwd{} is too conservative, we lose flexibility to get even better allocation in decision-making.
In \S\ref{sec:hpt-exp-res} we  evaluate the influence of \ttfwd{}  with real experiments.
In practice, \ttfwd{} is not predictable because of the uncertainty of job submission to the main queue.
For a new system, however, we can look into the scheduler logs to extract a representative \ttfwd{} statistically and use that value in the optimization.
If \ttfwd{} shows high variance in real scheduler logs, an estimation (with reduced variance) based on the current state of scheduler queue using machine learning or advanced statistical techniques may benefit the optimization.

\subsection{Full Model}
The complete mixed-integer optimization problem is then
\[ \begin{array}{lll}
   \dps \maxi_{x} \; & \eqref{eq:objf} \quad   & \text{objective} \\
   \st  \;                    & \eqref{eq:cnt-jsz} \quad & \text{job size constraints} \\
                              & \eqref{eq:nodes-constraint} \quad & \text{resource allocation} \\
                              & \eqref{eq:xor} \eqref{eq:migration-constraint} \quad & \text{job migration constraints} ,
  \end{array}
\]
where we  list here only the variables $x$ that represents the resource allocation map. 
In actual runs, all auxiliary variables (e.g., $y^l$, $y^u$, $z^u$, $z^d, w$) are also resolved.
The MILP is implemented using Gurobi~\cite{glockner2015parallel} optimizer and open source \footnote{https://github.com/lzhengchun/BFTrainer}.

\subsection{MILP Solution Time Benchmark}
Solving a MILP problem is NP-hard.
We need to solve a MILP problem,
with varying size and values of constant variables,
each time \pool{} or ${\mathcal J}$ change, in order to optimally reallocate nodes for \bfjobs{}.
If solving the MILP problem for a particular event takes time $t$, then idle nodes are used suboptimally for $t$ seconds. 
For example, if $n$ nodes join \pool{} at time $t_0$, these $n$ nodes will be idle until $t_0+t$.
Thus, the time to solve the MILP problem affects resource utilization efficiency. 
Here, we benchmarked the time to solve the MILP for varying ${\mathcal J}$ and ${\mathcal N}$,
using the Gurobi~\cite{glockner2015parallel} optimizer on a commodity computer without any customized improvement schemes such as heuristic branch-and-cut algorithms.
Since the time to solve a MILP problem also depends on the initial condition, we repeated each experiment 10 times with different initial conditions and took an average.  
While one can optimize solving MILP problems in many ways, exploring these methods is out of the scope for this study.

\begin{figure}[htb]
\vspace{-.3cm}
\center
\includegraphics[width=\columnwidth, trim=5cm 0 0 2cm, clip ]{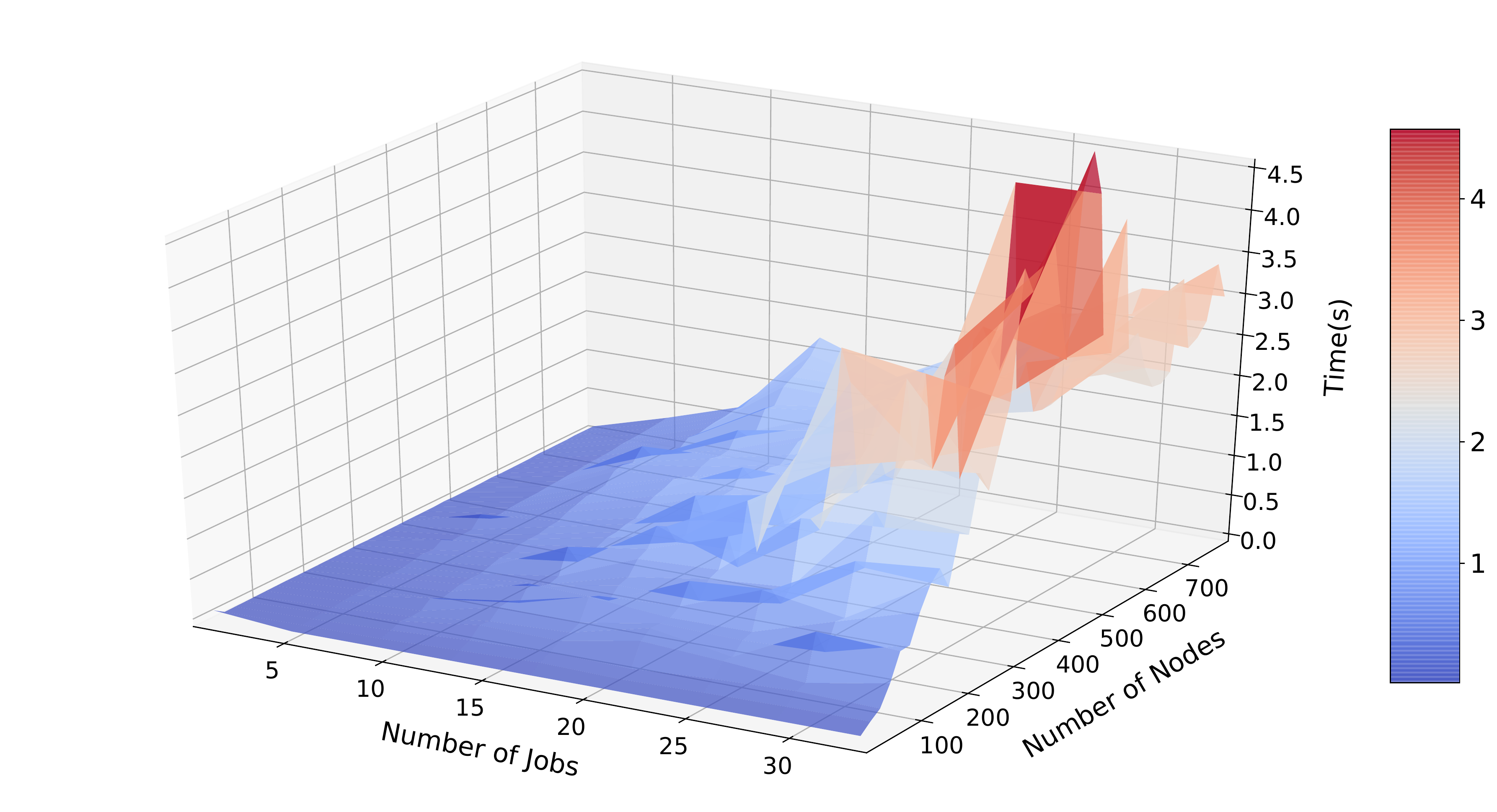} 
\caption{Optimization time vs.\ number of jobs and nodes.}
\label{fig:mip-time-bm}
\vspace{-.4cm}
\end{figure}

Although \pool{} occasionally exceeds more than one thousand nodes, this occurs only rarely and briefly, as shown in~\autoref{fig:spans-cdf}.
The \textit{average} number of idle nodes is about 9\% of of Summit: about \num{400} nodes. 
Thus we set the maximum MILP problem size in this benchmark to be \num{800} nodes.
We randomly initialized $c$ to meet all constraints for each repetition, called the MILP solver to reallocate nodes for \bfjob{}s, and measured the wall time taken. 
The solution time on a 2.3~GHz 8-Core Intel~Core~i9 processor (see  
\autoref{fig:mip-time-bm}) is typically less than one second:
negligible because relatively short when compared with the time gaps between events. Furthermore, nodes in \pool{} are not idle but just used suboptimally during the optimization.

Since all variables except the S2 are binary,
the high dimensionality of the model is the major contributor to computation costs.
The resulting search problem thus has limited opportunities for parallelism, 
and the best approach to further acceleration is likely to be
using processors with better single-core performance~\cite{koch2011miplib}.
Systems with higher clock rates (instead of more cores) and the fastest available memory can be used with parallel and distributed optimization solvers~\cite{glockner2015parallel} to accelerate solving large MILP problems.
Furthermore, because MILP problems are generally solved by using a LP-based branch-and-bound algorithm, we can set a timeout for the MILP solver in practice.
If the solver returns a status of timeout but finds feasible solutions, we pick the better one between the current map $c$ and the best-so-far solution.
Otherwise, if it does not come back with any feasible solution within the time limit, we keep the current nodes-to-\bfjob{}s map (i.e., let $x$ be $c$).

\section{Experimental Setup}\label{sec:exp-setup}
We introduce evaluation metrics, computing resources, DNN training scenarios, and design of the experiment for evaluating resource utilization and observing best practices of parameter configuration.

\subsection{Evaluation Metrics}\label{sec:metrics}
\subsubsection{Resource integral}
Makespan, a widely used metric for evaluating scheduling policies, is not well suited to our context because of the frequent fluctuations in \pool{}. 
For example, if we observe on a computer system of fixed size that two policies take 10 and 11 hours, respectively, to run the same workload, we may conclude that the second policy is just 10\% less efficient than the first. 
In our context, however, there may have been many more idle nodes during that 11th hour. 
Thus, we instead use the actual resources consumed---the integral of $|$\pool{}$|$ over time: \autoref{eq:machinetime}---for comparing different policies. 
It is the same as makespan times the number of nodes (i.e., leads to \mt{}) in conventional scheduling for dedicated nodes.
Specifically,  assume that there are $K$ events $\{e_1, e_2, \ldots e_K\} \in \mathcal{E}$ during a given period of $t=T_{e_K} - T_{e_1}$ seconds.  
Then the total resource amount can be quantified, in \nodehour{}, as
\begin{equation}
R_{e_1 \to e_K}=\sum_{k=1}^{K-1}N_{e_k}\frac{T_{e_{k+1}}-T_{e_k}}{3600},
\label{eq:machinetime}
\end{equation}
where $N_e$ is the size of \pool{} between $e_k$  and $e_{k+1}$.
This amount is theoretical; the effective resources usable for running \bfjob{}s will be smaller than $R_{e_1 \to e_K}$ because both preemption (when nodes leave \pool{}) and rescaling \bfjob{}s (when $\mathcal{J}$ or \pool{} change) incur costs.

\subsubsection{Resource utilization efficiency}

$R_{e_1 \to e_K}$ is equivalent to having a static $N_{e_1 \to e_K}^{eq}$ nodes for $t=T_{e_K} - T_{e_1}$ seconds from the perspective of \nodehour{}, where
\begin{equation}
N_{e_1 \to e_K}^{eq} = \frac{\sum_{k=1}^{K-1}N_{e_k}(T_{e_{k+1}}-T_{e_k})}{T_{e_K} - T_{e_1}}.
\label{eq:eqnds}
\end{equation}

Let $A_e$ be the total outcome (e.g., total samples processed) when \proj{} is used to run a set of \bfjob{}s, $\mathcal{J}$, from $T_{e_1}$ to $T_{e_K}$,
and $A_s$ the total outcome if we run $\mathcal{J}$ using static $N_{e_1 \to e_K}^{eq}$ nodes for $t$ seconds (assume $\mathcal{J}$ is too large to finish in $t$ seconds).
Since there is no additional cost when using static $N_{e_1 \to e_K}^{eq}$ nodes, we can use $A_s$ as a baseline for the machine time of $R_{e_1 \to e_K}$ to run the given list of \bfjob{}s using \proj{} for $t$ seconds. 
We can then define the resource utilization efficiency of \proj{} as $U=\frac{A_e}{A_s}*100\%$.
While the unavoidable preemption costs incurred when nodes leave \pool{} mean that $A_e<A_s$, we can improve $A_e$ by identifying a better resource allocation for \bfjob{}s (e.g., via the proposed MILP model).

\subsection{DNN Training Scenarios}
Data parallelism is commonly used for DNN training in distributed-computing environments~\cite{ben2019demystifying}.
In data parallelism, a model is replicated across a set of distributed nodes, and each minibatch is divided into equal-sized partitions per node.
Each node computes a local gradient using its own data partition; these local gradients are then averaged across all replicas (typically by using a collective all-reduce operation or parameter servers) to obtain the desired gradients for the optimizer.
Then, on each node the optimizer updates the weights of the model replica using the desired gradients. 

In an isolated network environment such as HPC, the time needed to average local gradients and synchronize model parameters across all \bfjob{} replicas depends only on model size.
According to Amdahl's law~\cite{Amdahl67}, the runtime for each training iteration is lower bounded by the model synchronization time.
Therefore, it is desirable to increase the batch size so that a larger proportion of time can be spent on computing the local gradient estimates instead of synchronizing gradients and the model~\cite{qiao2020pollux}.
However, although a larger batch size is beneficial to computation, simply increasing batch size without also increasing learning rate cannot accelerate model training convergence~\cite{smith2017don,you2018imagenet}. 

The learning rate and batch size are training parameters typically chosen by the model developer.
In \proj{}, the per-node minibatch size is fixed, and thus the global batch size depends on the number of nodes that \proj{} allocates to a \bfjob{}. 
The user specifies on submission the minimum and maximum number of nodes on which the application can run; and a learning rate scheduler that can adjust learning rate according to the global batch size~\cite{goyal2017accurate,you2018imagenet}.
\autoref{tbl:apps-benchmark} shows the (weak) scaling of several DNN models for the ImageNet dataset on Summit using a minibatch of 32 per GPU.
Of these seven DNNs, AlexNet, developed in 2012, is the least compute-intensive (727 MFLOPs/sample), and DenseNet, published in 2016, is the most compute-intensive (3 GFLOPs/sample).
We ran 100 epochs for each DNN in \autoref{tbl:apps-benchmark}, corresponding to processing 130 million samples in each case.
We note that although we consider weak scaling in this experiment, a \proj{} user can employ either strong scaling, with a static global batch size, or weak scaling, with strategies to deal with dynamic global batch size~\cite{maleki2021scaling}.
To make the evaluation more representative, we designed experiments that need to run at least one week in practice.

\begin{table}[htb]
\vspace{-2mm}
\centering
\caption{Performance of ImageNet models, in samples per second ($\times$1000), vs.\ number of nodes when running with data parallelism and a minibatch of 32 per GPU on Summit.}
\begin{tabular}{l|c|r|r|r|r|r|r}
\noalign{\hrule height 2pt}
\backslashbox{\bf DNN}{\bf Nodes} & \multicolumn{1}{c|}{1} & \multicolumn{1}{c|}{2} & \multicolumn{1}{c|}{4} & \multicolumn{1}{c|}{8} & \multicolumn{1}{c|}{16} & \multicolumn{1}{c|}{32} & \multicolumn{1}{c}{64} \\\hline\hline
AlexNet~\cite{alexnet12}       & 7.1 & 13.1 & 21.1 & 40.5 & 74.0 & 130.8 & 202.1 \\
ResNet18~\cite{resnet16}       & 5.2 & 10.6 & 20.4 & 39.6 & 78.0 & 144.8 & 262.7 \\
MnasNet~\cite{Mnasnet}         & 3.2 & 6.0 & 11.5 & 23.1 & 43.9 & 83.5 & 160.5 \\
MobileNets~\cite{mobilenets17} & 3.0 & 5.9 & 11.4 & 22.0 & 42.5 & 82.3 & 155.2 \\
ShuffleNet~\cite{shufflenet18} & 2.8 & 5.3 & 10.0 & 20.4 & 38.9 & 74.1 & 145.1 \\
VGG-16~\cite{vgg14}            & 1.2 & 2.4 & 4.7 & 9.3 & 18.3 & 36.2 & 70.2 \\
DenseNet~\cite{densenet16}     & 1.0 & 2.0 & 3.8 & 7.6 & 15.0 & 28.8 & 57.8 \\
\noalign{\hrule height 2pt}
\end{tabular}
\label{tbl:apps-benchmark}
\vspace{-0.4cm}
\end{table}


\subsection{Use of Summit Logs to Drive Simulations}
\begin{figure}[htb]
\centering
\includegraphics[width=0.8\columnwidth]{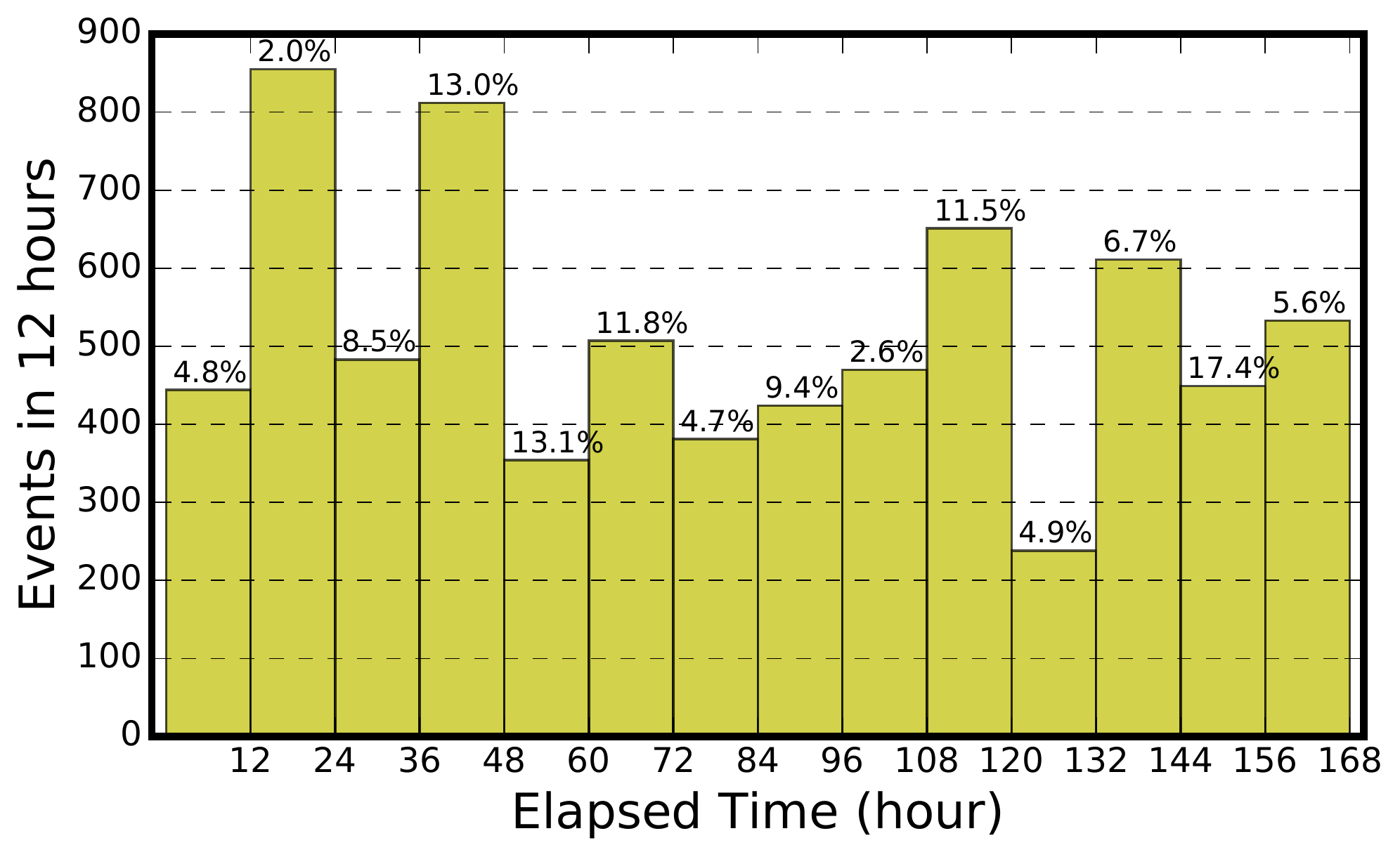}
\caption{Characteristics of idle nodes, \pool{}, over a week. The number above each bar indicates the percent of all nodes that are idle.}
\label{fig:week-idle-nodes-character}
\vspace{-.4cm}
\end{figure}

We used events in logs for an arbitrarily chosen 1024 Summit nodes over a week (see \autoref{fig:week-idle-nodes-character}) to drive our simulations of \pool{} behavior.
We used Horovod~\cite{sergeev2018horovod} with PyTorch~\cite{paszke2019pytorch} and model architecture implementations from the torchvision package (a part of the PyTorch project) for the experiments in this study. 
Elastic Horovod enables Horovod to scale up and down the number of workers (i.e., nodes in this study) dynamically at runtime, without requiring a restart or resuming from checkpoints saved to durable storage.
To use \proj{}, a user simply provides a training script as  would be done if   running in a conventional manner, with the one difference being that it needs to have Elastic Horovod enabled. 
\proj{} then runs the resulting \bfjob{} elastically, starting it or rescaling it to fit in a series of suitable fragments whenever the MILP algorithm indicates that it is cost-effective to do so.
\proj{} can also work with other DNN distributed training frameworks such as AdaptDL~\cite{qiao2020pollux} and TorchElastic (a framework of PyTorch~\cite{paszke2019pytorch}).
In principle, any iterative applications that can scale up and down at runtime with reasonably small cost can be run using \proj{}.

\section{Experiments and Evaluations}\label{sec:exp-ana}
We present experiment results for \proj{} in two scenarios: single-user HPO/NAS and as a central resource manager for multiple users. 
Our experiments evaluate MILP in these different scenarios and provide guidance for best practices.

\subsection{Hyperparameter Optimization}\label{sec:hpt-exp-res}
HPO is an essential step in deep learning workflows.
It increases model quality by systematically evaluating many different sets of hyperparameters (each called a ``trial") and picking the best outcome.
HPO is resource intensive, with the resource demand roughly proportional to the number of trials performed.

In the HPO context, we assume that all trials within an experiment have the same scaling properties (parameters that control  the model architecture are considered as NAS)~\cite{hypersched}.
We arbitrarily chose ShuffleNet~\cite{shufflenet18} from \autoref{tbl:apps-benchmark} to evaluate \proj{} for HPO.
We used total throughput from all nodes as the objective metric to optimize when allocating nodes to \bfjob{}s.
ShuffleNet is a high-accuracy neural network with 5.4~M parameters and 524~M multiply-accumulates (MACs); its relatively low number of FLOPS makes it suitable for edge devices, and it usually benefits more from HPO when dealing with the trade-off between MACs and accuracy~\cite{heydarigorji2020hypertune}.
In this experiment, we ran \num{1000} trials while replaying Summit logs for the dynamics of \pool{}. \autoref{fig:week-idle-nodes-character} shows the statistics for the first 168 hours.
Depending on the parameter \ttfwd{} in MILP, about 200 hours of log time are needed to complete all \num{1000} trials.

We compare these results against those achieved with a baseline scheme that distributes nodes equally to \bfjob{}s.
If there is no preemption within the forward-looking time, MILP is guaranteed to perform no worse than the baseline because the heuristic also meets all MILP constraints. 
That is, the baseline scheme produces the optimal MILP solution when there is no rescaling cost and no preemption (i.e., resources are dedicated).
We also conducted experiments with different \ttfwd{} to study the influence of \ttfwd{} and identify guidelines for picking a good \ttfwd{}.

\subsubsection{Forward-looking time implication}\label{sec:tfwd}
\begin{figure}[ht]
\vspace{-.3cm}
\centering
\subfloat[Preemption within \ttfwd{}.]{\label{fig:hpt-fwd-vs-preempt-within-fwd}{\includegraphics[width=0.5\linewidth]{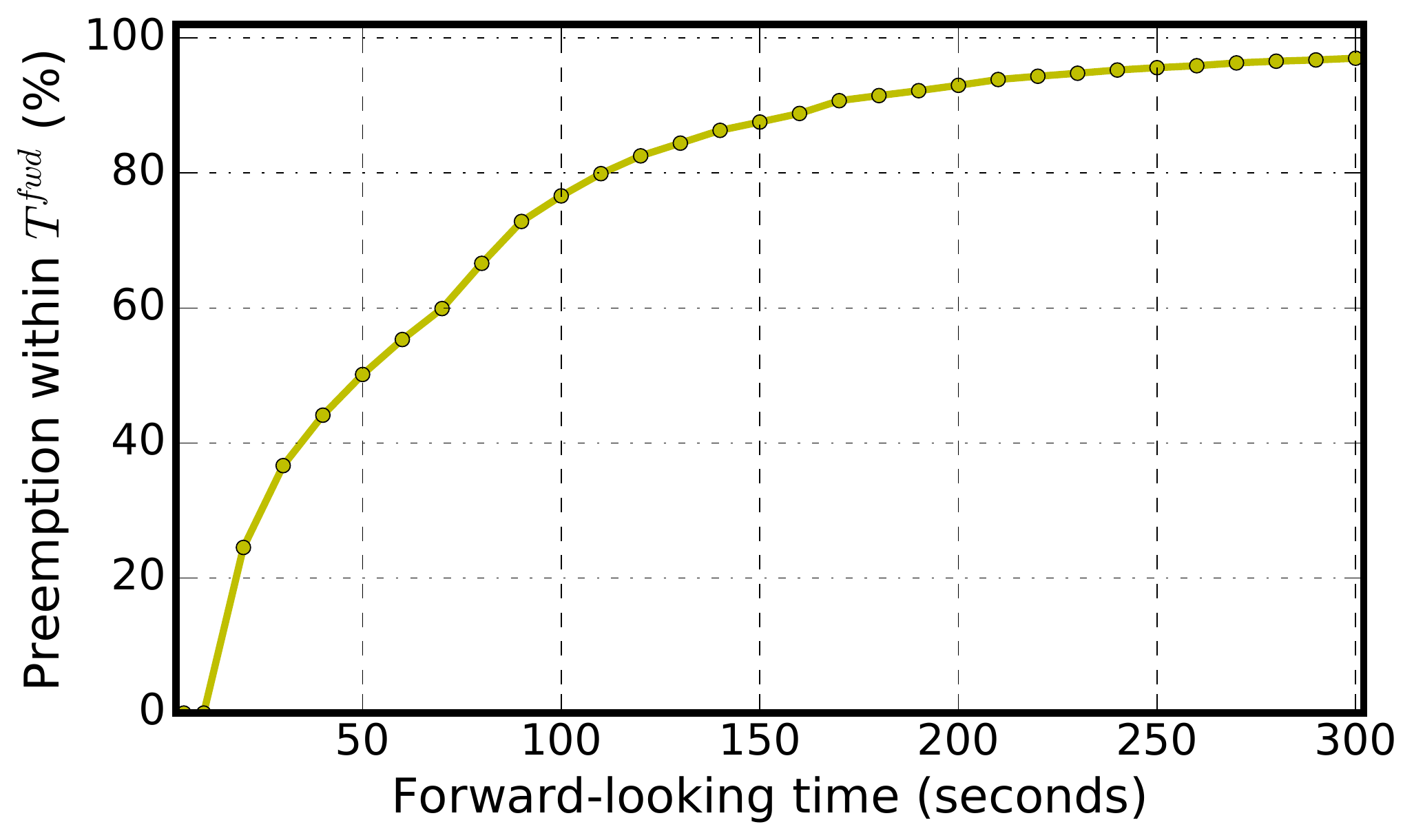}}}
\subfloat[Rescaling cost per event.]{\label{fig:hpt-fwd-vs-rescale-cost}{\includegraphics[width=0.5\linewidth]{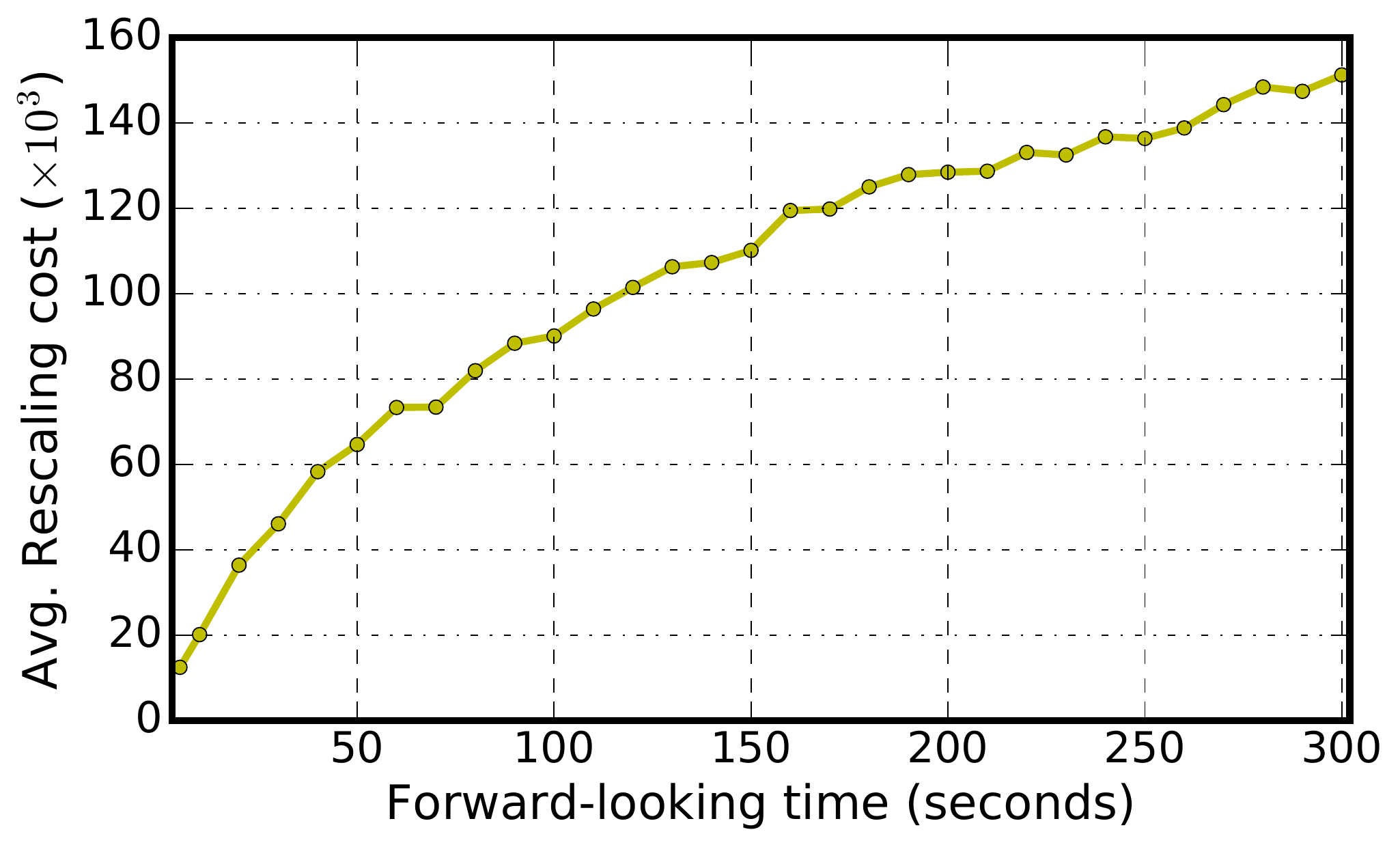}}}
\caption{Influence of forward-looking time on preemption.}
\label{fig:hpt-fwd}
\vspace{-.4cm}
\end{figure}

\autoref{fig:hpt-fwd-vs-preempt-within-fwd} shows the percentage of preemption within the forward-looking time.
As expected, the chance of preemption within forward-looking time increases with \ttfwd{} and reaches 90\% when \ttfwd{} $\ge 170$ seconds.

Since the MILP tries to maximize total outcome and minimize rescaling cost (i.e., \autoref{eq:objf}), we show in \autoref{fig:hpt-fwd-vs-rescale-cost} the rescaling cost.
This cost increases with forward-looking time because the rescaling cost is static regardless of the forward-looking time.
The MILP model will try to aggressively rescale \bfjob{}s to maximize the aggregated outcome when the forward-looking time is long, because the rescaling cost is then small relative to total outcome in \ttfwd{}.
As a comparison, the average rescaling cost of the baseline is 1.03$\times$10$^6$ samples per event, that is, 76$\times$ more (worse) than the MILP with a forward-looking time of 10 seconds.

\begin{figure}
\vspace{-.3cm}
\centering
\includegraphics[width=\linewidth]{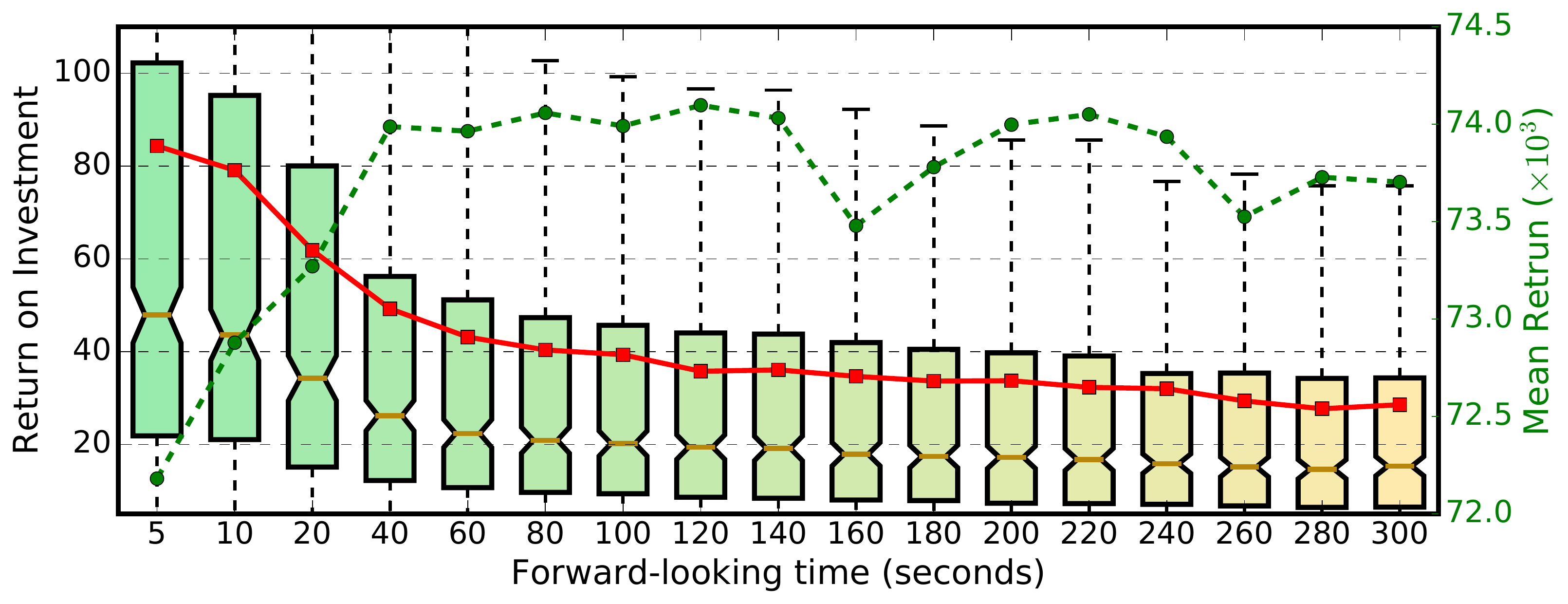}
\caption{Influence of forward-looking time on rescaling investment and return. Solid red line is average return on investment.}
\label{fig:hpt-roi}
\vspace{-.4cm}
\end{figure}

For a given event $e_i$, MILP optimally (re)allocates nodes to \bfjob{}s to maximize the total outcome within the given forward-looking time. 
If we view the rescaling cost as an investment and the total outcome from $e_i$ to $e_{i+1}$ as return, \autoref{fig:hpt-roi} compares the return on investment (ROI) when different forward-looking times are used.
Here we do not compare the return within the forward-looking time because it is an expectation. 
Preemption may happen within the forward-looking time, thus leading to a suboptimal operation. 
As one can see from \autoref{fig:hpt-roi},  the ROI decreases with the increase of forward-looking time. 
This decrease is because the rescaling cost (i.e., investment) increases with forward-looking time as shown in \autoref{fig:hpt-fwd-vs-rescale-cost} but the total outcome increases with forward-looking time only to some extent and then is saturated (green dotted line in \autoref{fig:hpt-roi}).
Moreover, since \ttfwd{} does not appear in any of MILP's constraints, the optimal solution for a shorter \ttfwd{} is a (suboptimal) solution of a longer \ttfwd{}, and vice versa.
Although a longer \ttfwd{} cannot keep increasing the throughput after a certain value, it does give a smaller yet more stable ROI.

\subsubsection{Resource utilization efficiency}
The dynamic nature of \pool{} 
introduces costs to \bfjob{}s as \proj{} needs to rescale them  to fit \pool{}.
We use resource utilization efficiency, $U$, defined in \S\ref{sec:metrics} to quantify how well MILP can use nodes in \pool{}. 
Basically three factors influence resource utilization efficiency: preemption cost, rescaling cost, and the scale at which \bfjob{}s run.
The preemption cost is caused by the dynamics of \pool{} and is outside of our control. 
Enabling \bfjob{}s to run at a more efficient scale needs more investment to rescale running \bfjob{}s, for example, by migrating nodes from large \bfjob{}s to smaller \bfjob{}s.
In our MILP, a longer \ttfwd{} allows for more efficient resource allocation, since the MILP can then expect a better return and can thus invest more in rescaling.

\begin{figure}[htb]
\vspace{-.3cm}
\centering
\includegraphics[width=0.8\columnwidth]{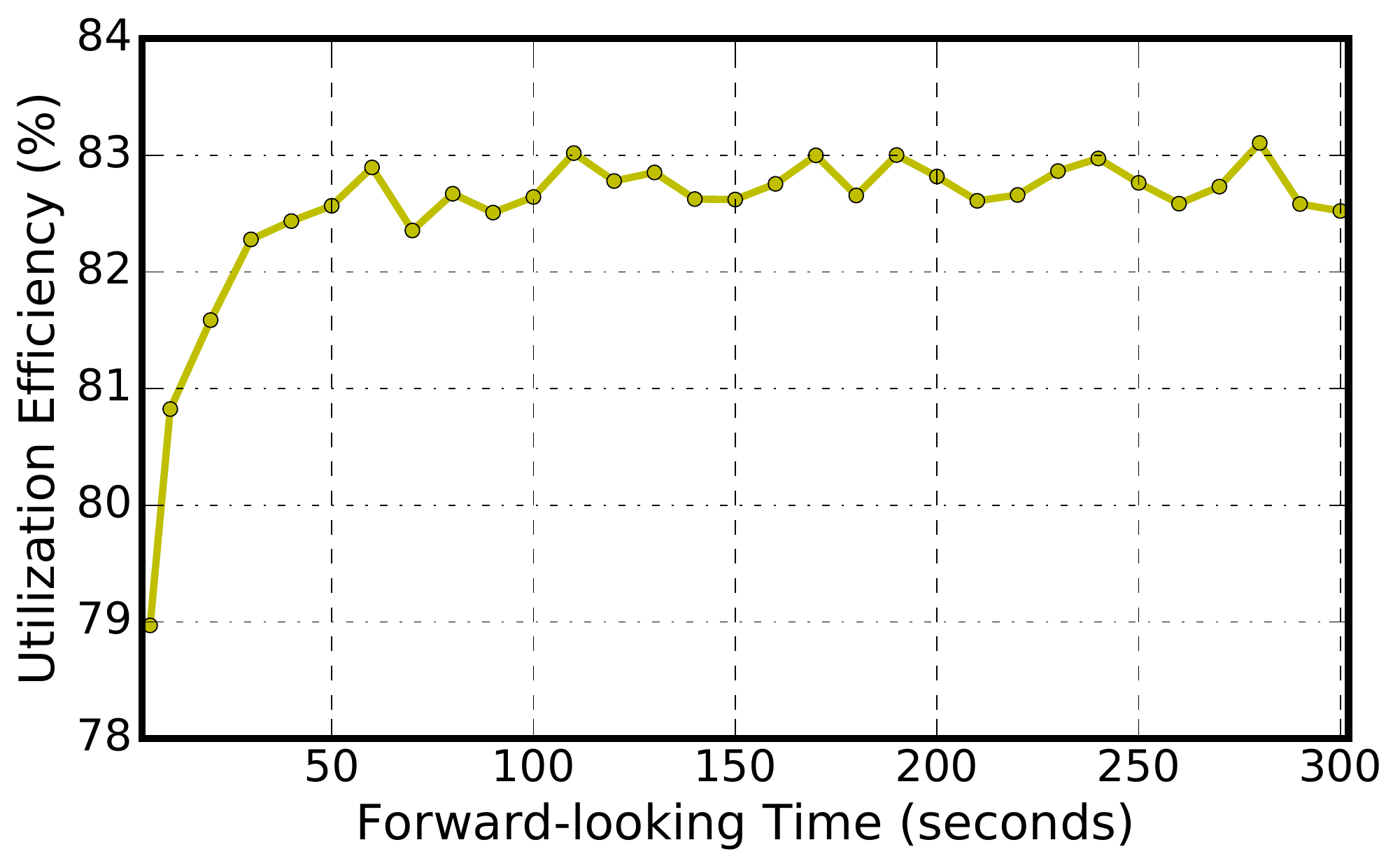}
\caption{HPO resource utilization efficiency as a function of forward-looking time.}
\label{fig:hpt-fwd-vs-effi}
\vspace{-.4cm}
\end{figure}

\autoref{fig:hpt-fwd-vs-effi} shows how resource utilization efficiency varies with \ttfwd{}.
Efficiency initially increases with \ttfwd{} but soon saturates at around \ttfwd{}=120 seconds. We note that the efficiency of the heuristic scheme is 75\%.
As shown in \autoref{fig:hpt-fwd-vs-rescale-cost}, although a longer \ttfwd{} results in more rescaling cost, the resource utilization is not affected, as seen in \autoref{fig:hpt-fwd-vs-effi}.
The reason is that there are two ways to achieve high efficiency: (1) save cost by using a conservative \ttfwd{} but limit rescaling \bfjob{}s to run at the most efficiency scale and (2) invest more cost to enable rescaling \bfjob{}s to run at more efficient scale.
In  some cases both ways can lead to similar efficiency based on the dynamics of \pool{} (e.g., the one we used in this experiment).

\begin{figure}[htb]
\vspace{-.3cm}
\centering
\includegraphics[width=0.8\columnwidth]{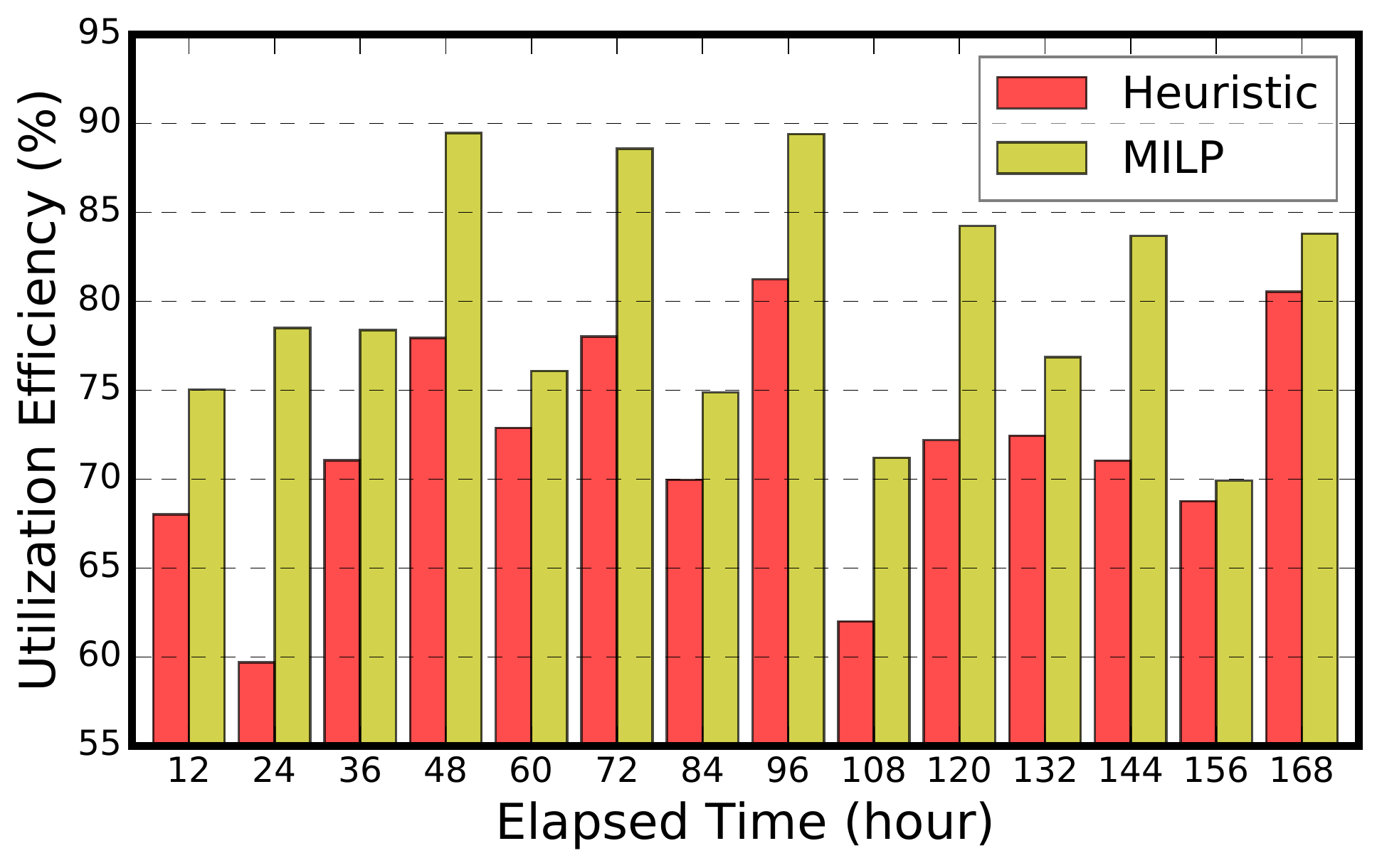}
\caption{HPO resource utilization efficiency over a week.}
\label{fig:hpt-efficiency}
\vspace{-.4cm}
\end{figure}

To get more insight into the relationships between the dynamics of \pool{} and $U$, we focused on \ttfwd{}=120 seconds and computed $U$  every six hours in a week. 
\autoref{fig:hpt-efficiency} shows results for the MILP and heuristic schemes.  
As expected, MILP performs better than the heuristic, achieving
up to 32\% higher efficiency.
On average, MILP achieved an efficiency of 80\% and up to $\sim$90\% for three of the six-hour periods. 
The efficiency also depends on the size of \pool{} and on dynamic features such as the number of events, as shown in \autoref{fig:week-idle-nodes-character}. 
For example, if nodes join and leave \pool{} only rarely, both the heuristic and MILP should easily get close to 100\%.
However, if events are frequent, that is, if the length of fragments is extremely short, it is impractical to get high efficiency because of the unavoidable rescaling cost and preemption.

\begin{figure}[ht]
\vspace{-.3cm}
\centering
\subfloat[Jobs preemption cost.]{\label{fig:hpt-preemption-cost}{\includegraphics[width=0.5\linewidth]{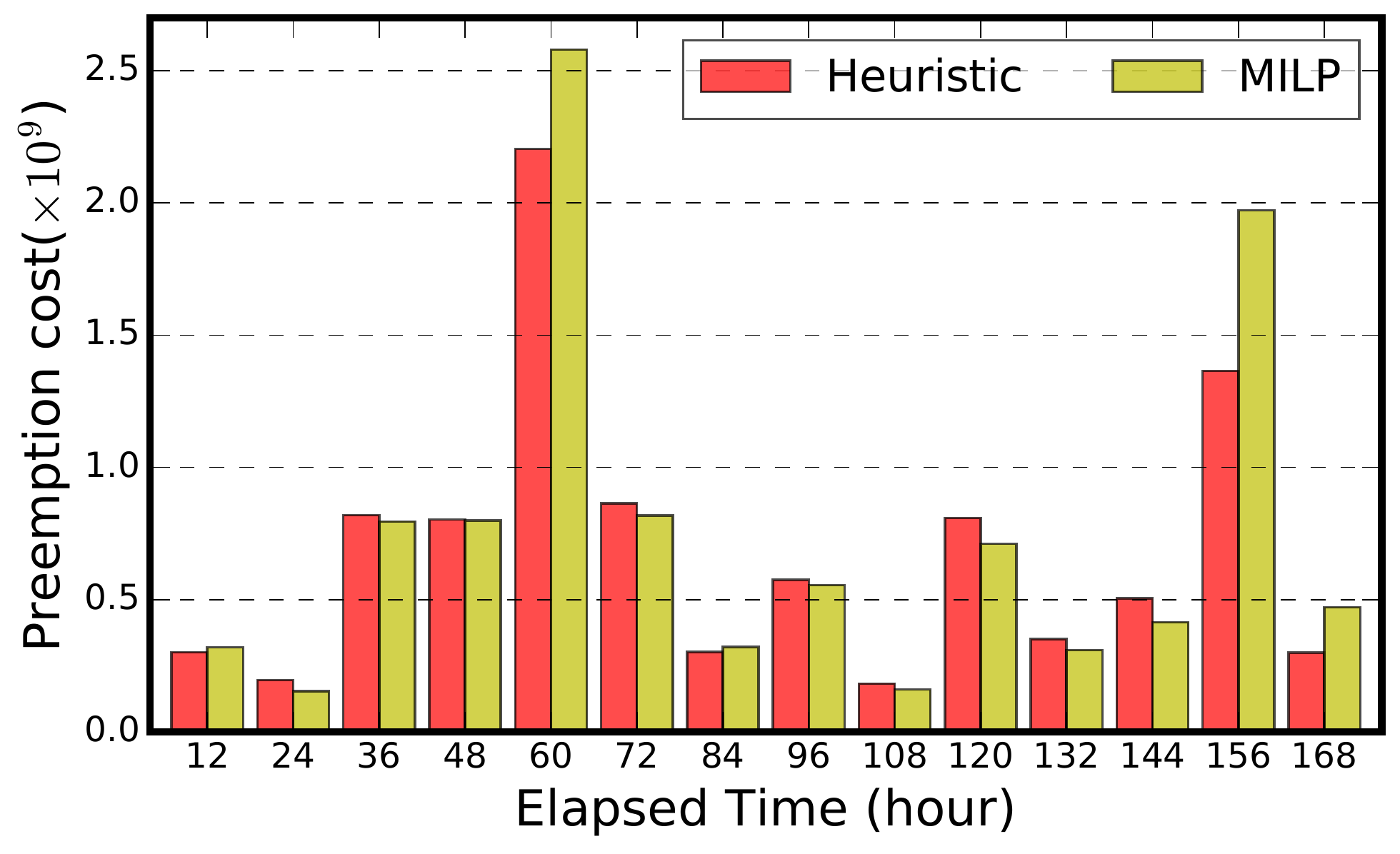}}}
\subfloat[Rescaling cost.]{\label{fig:hpt-log10rescost}{\includegraphics[width=0.5\linewidth]{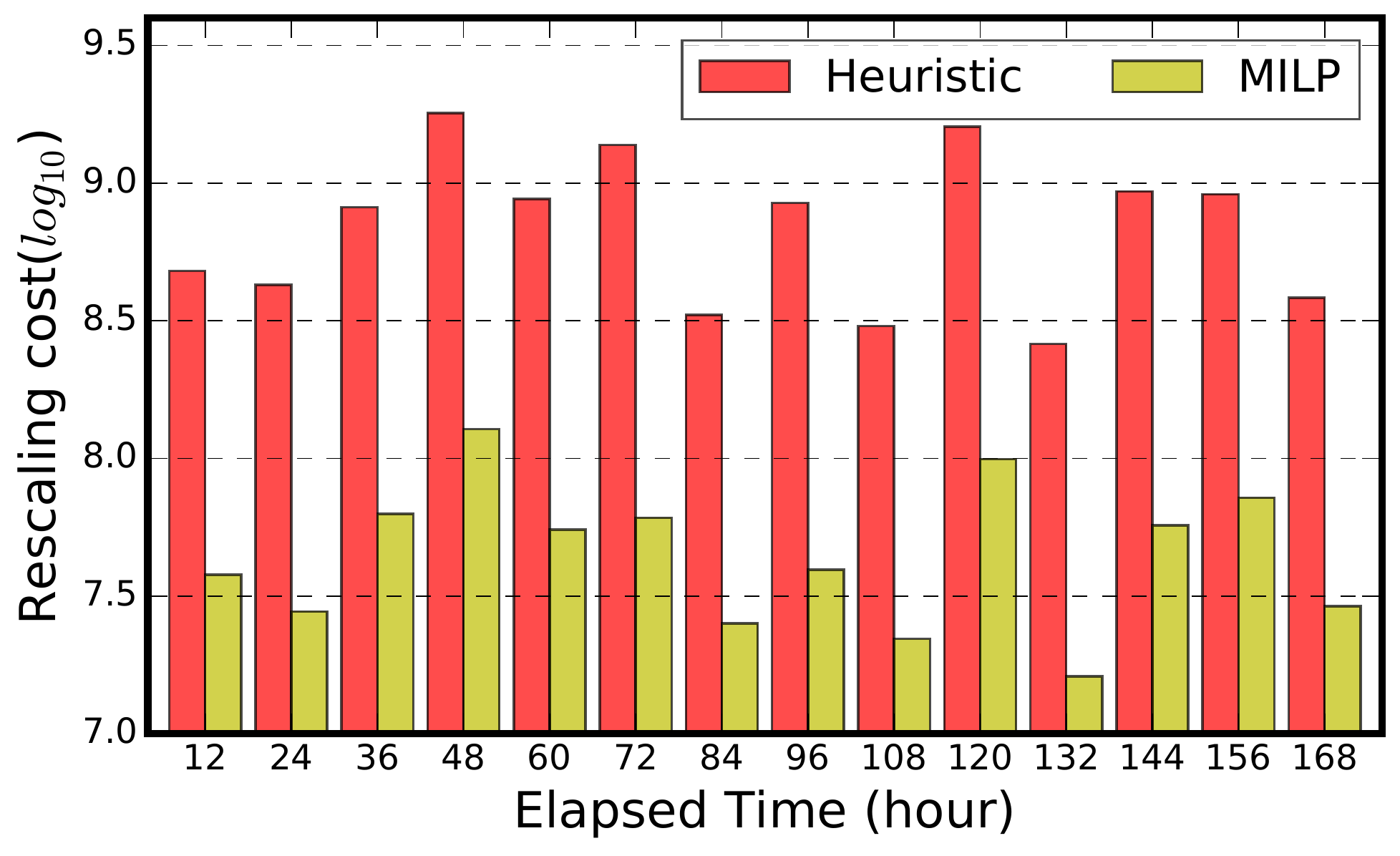}}}
\caption{Job preemption and rescaling costs over a week.}
\label{fig:hpt-cost}
\vspace{-.4cm}
\end{figure}

To obtain further insight into why MILP can use fragmented resources efficiently, we computed its preemption and rescaling costs (\autoref{fig:hpt-cost}). 
We ee do not control the preemption cost we see in \autoref{fig:hpt-preemption-cost} little difference between the heuristic and MILP.
Since MILP tries to maximize aggregated throughput (within the forward-looking time), thus also minimizing the rescaling cost in \autoref{eq:objf}, its rescaling cost is much less than that of the heuristic (\autoref{fig:hpt-log10rescost}).
This explains why MILP uses resources more efficiently (\autoref{fig:hpt-efficiency}).

To evaluate the effectiveness of MILP resource allocations, we computed the number of images processed between pairs of contiguous events and 
divided the outcome of \proj{} by that of the baseline scheme to get a speedup metric. 
We find that because the change of preemption is unknown, MILP is not always superior to the heuristic,
performing worse for 6.9\% of events.
However, rescaling cost and \bfjob{} scalability considerations lead to MILP being better than the heuristic in most cases: 2$\times$ and 5$\times$ better for 11\% and 3\% of events, respectively. 
Average MILP speedup is 2.9$\times$, but we note that this is not a reliable metric in this context because it can be skewed by high speedup of a few events; the total outcome also depends on the elapsed time between contiguous events.
\begin{obsbox}[boxsep=1pt,left=8pt,right=2pt,top=0pt,bottom=0pt]{green}{green}
\obs{HPO is a good application scenario for \proj{} because it is resource-consuming and all trials have the same scalability.  
The \tfwd{} is a parameter of the MILP that needs to be set based on the dynamics of the \pool{}.
A small \ttfwd{} discourages rescaling \bfjob{} that leads to inefficient use of fragments.
The MILP is more robust to a large \ttfwd{} setting because it relaxes the rescaling cost and thus enables exploring more efficient use of fragments.
}
\end{obsbox}

\subsection{Diverse \bfjob{}s}\label{sec:diverse-job}
We evaluated a scenario in which \bfjob{}s in $\mathcal{J}$ are diverse in order to mimic the scenario where  only one \proj{} instance is managed by administrators and users submit DNNs with diverse scalability.
It can also mimic scenarios in which a user employs \proj{} for NAS.
We modeled \bfjob{} submission as a Poisson process but limited the number of concurrent \bfjob{}s to 10 in this experiment. 
The characteristics of the DNN for each \bfjob{} submission were cyclically sampled from \autoref{tbl:apps-benchmark} for \num{1000} \bfjob{}s.

\subsubsection{Objective metric}
Resource allocation aims to rescale \bfjob{}s to optimize the aggregated objective metric. 
Besides constraints such as the minimum and maximum scale of each \bfjob{}, the main consideration when rescaling a \bfjob{} is the trade-off between rescaling cost and potential (based on the forward-looking time) increase in the objective metric.  
Throughput is a straightforward objective metric to maximize.
However, MILP biases resource allocation to high throughput rather than compute-intensive \bfjob{}s when scheduling \bfjob{}s with diverse scalability and resource demands.
Another metric can be scaling efficiency, a normalized version of throughput that
is agnostic to DNN throughput.

\begin{figure}[htb]
\vspace{-.3cm}
\centering
\includegraphics[width=0.8\columnwidth]{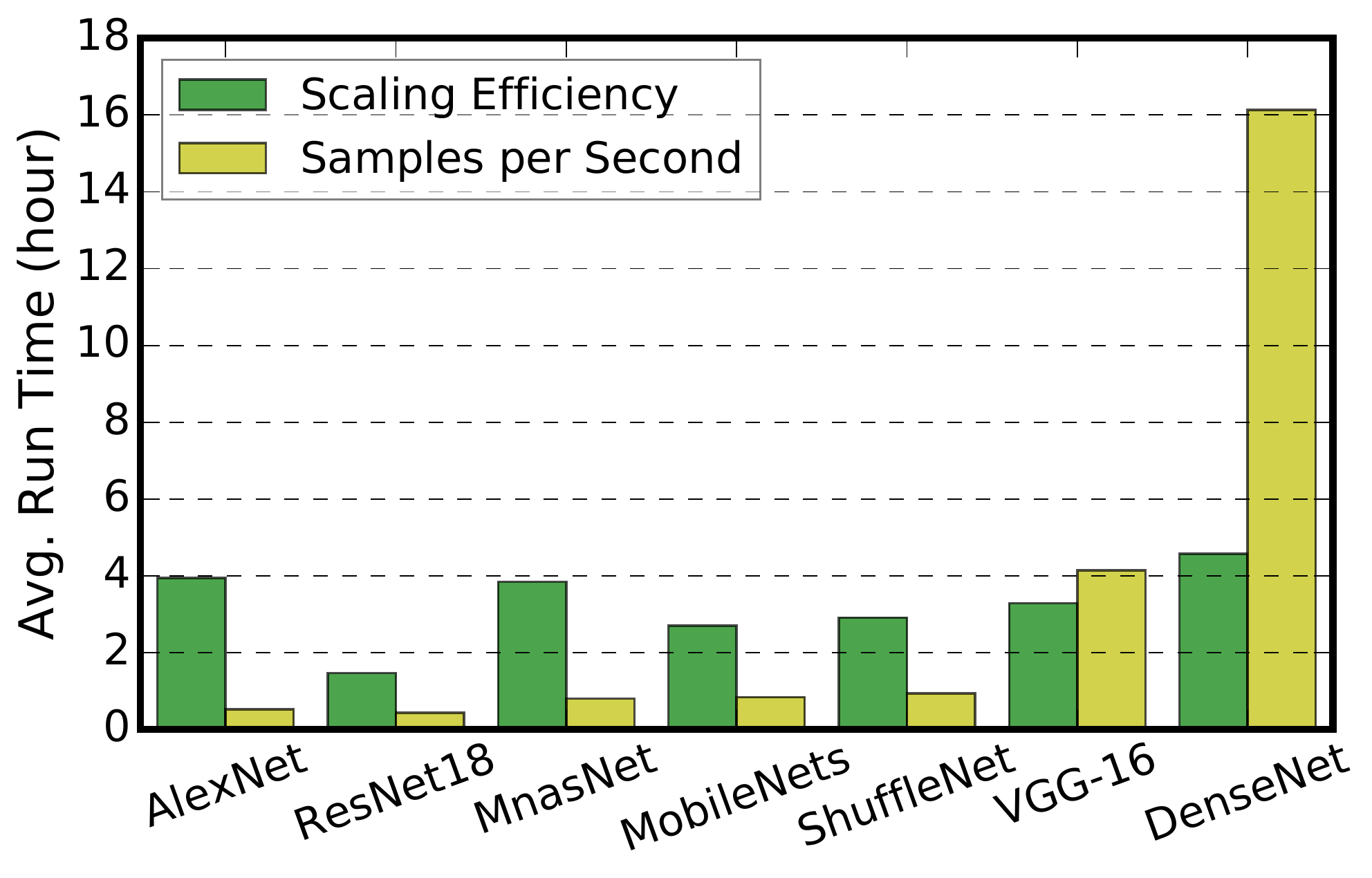}
\caption{DNN times when optimizing for two different objectives.}
\label{fig:spdup-vs-imgps-rt-maxJ15} 
\vspace{-.4cm}
\end{figure}

We used both metrics for a comparison study to elucidate their implications for recommending best practices.
\autoref{fig:spdup-vs-imgps-rt-maxJ15} compares the average runtime of different DNNs using the  different optimization metrics. 
We see that every DNN gets similar runtime when using the normalized scaling efficiency as the objective metric, although their computing demands are  different according to \autoref{tbl:apps-benchmark}.
When using throughput as the metric, however, MILP apparently gives priority to high-throughput DNNs such as AlexNet; compute-intensive DNNs such as DenseNet take much more time.
We note that although different DNNs will have different runtimes even if the same resources are allocated because of their difference in computing demand, it is not as different as  shown in \autoref{fig:spdup-vs-imgps-rt-maxJ15} according to \autoref{tbl:apps-benchmark}.
For example, the difference between Alexnet and DenseNet on throughput is only about 7$\times$, but the runtime difference is more than 40$\times$ on average and up to 70$\times$. 
We conclude that throughput is not a good metric in terms of fair share when \bfjob{}s differ significantly in their throughput.
We note that although MILP biases resource allocation to higher-throughput \bfjob{}s 
when raw throughput is used as the objective metric, it still needs to follow the FCFS policy.
Thus, limiting maximum parallel jobs helps alleviate starvation of \bfjob{}s with low throughput (see \S\ref{sec:maxj} for details).

\begin{figure}[htb]
\centering
\includegraphics[width=0.8\columnwidth]{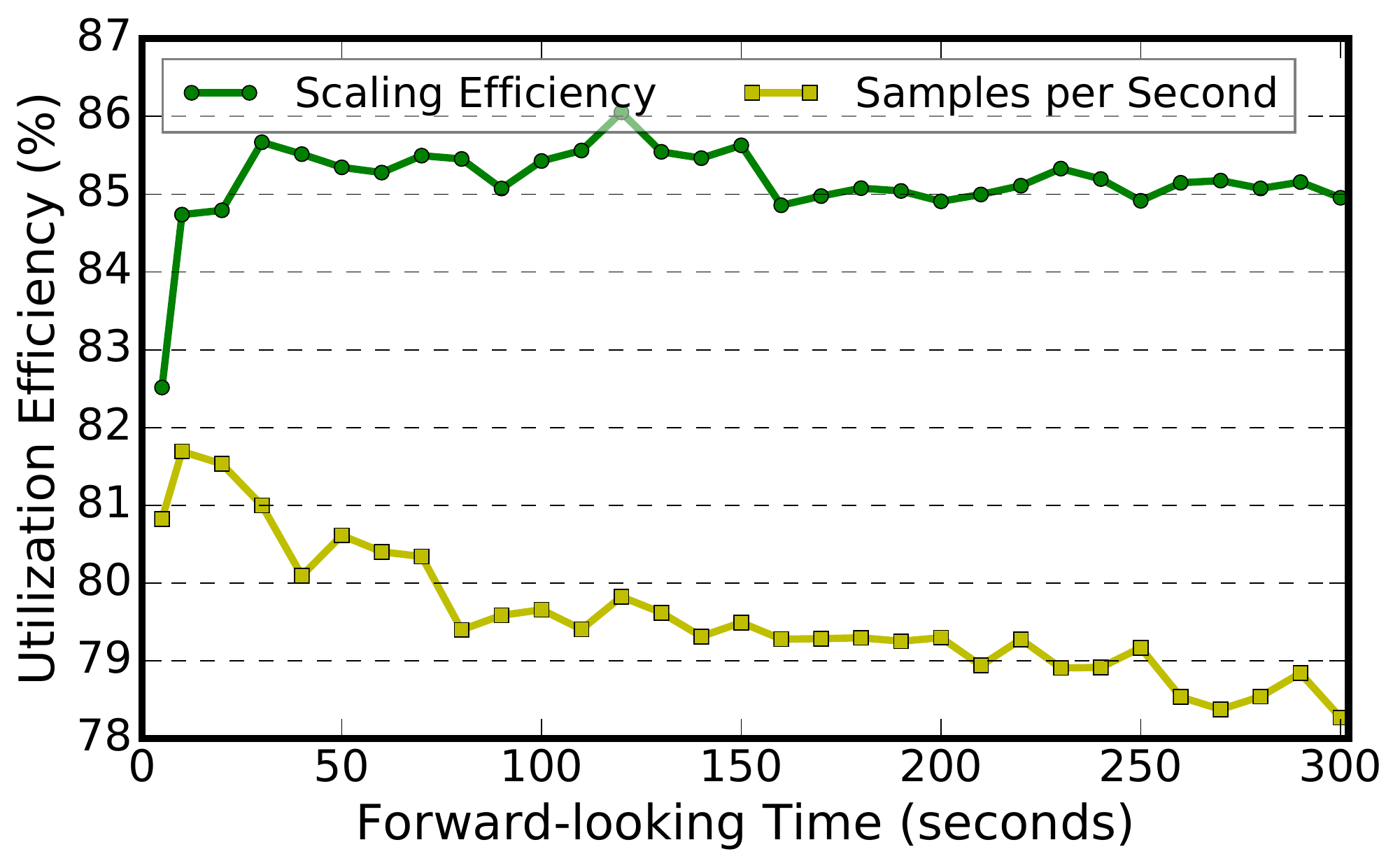}
\caption{Resource utilization efficiency using different objective metrics and \tfwd{}.}
\label{fig:spdup-vs-imgps-fwd-vs-effi}
\vspace{-.4cm}
\end{figure}

\autoref{fig:spdup-vs-imgps-fwd-vs-effi} shows how resource utilization efficiency varies with objective metric and \tfwd{}.
The influence of the latter on efficiency is similar to in the HPO scenarios.
$U$ is consistently better when scaling efficiency is used as the objective metric because MILP biases resources to high-throughput \bfjob{}s, leading them to run on many nodes but with reduced efficiency. 

\begin{obsbox}[boxsep=1pt,left=8pt,right=2pt,top=0pt,bottom=0pt]{green}{green}
\obs{\proj{} understands user goals through an objective metric.
Selecting this metric is straightforward for a single user training diverse DNNs (e.g., for NAS) because the user has a single goal.
With multiple users, the choice is a game
between resource provider and users; a \bfjob{}-wise normalized metric provides better fairness. 
}
\end{obsbox}

\subsection{Maximum Parallel \bfjob{}s}\label{sec:maxj}
\begin{figure}[ht]
\centering
\subfloat[]{\label{fig:pj-vs-mt}{\includegraphics[width=0.33\linewidth]{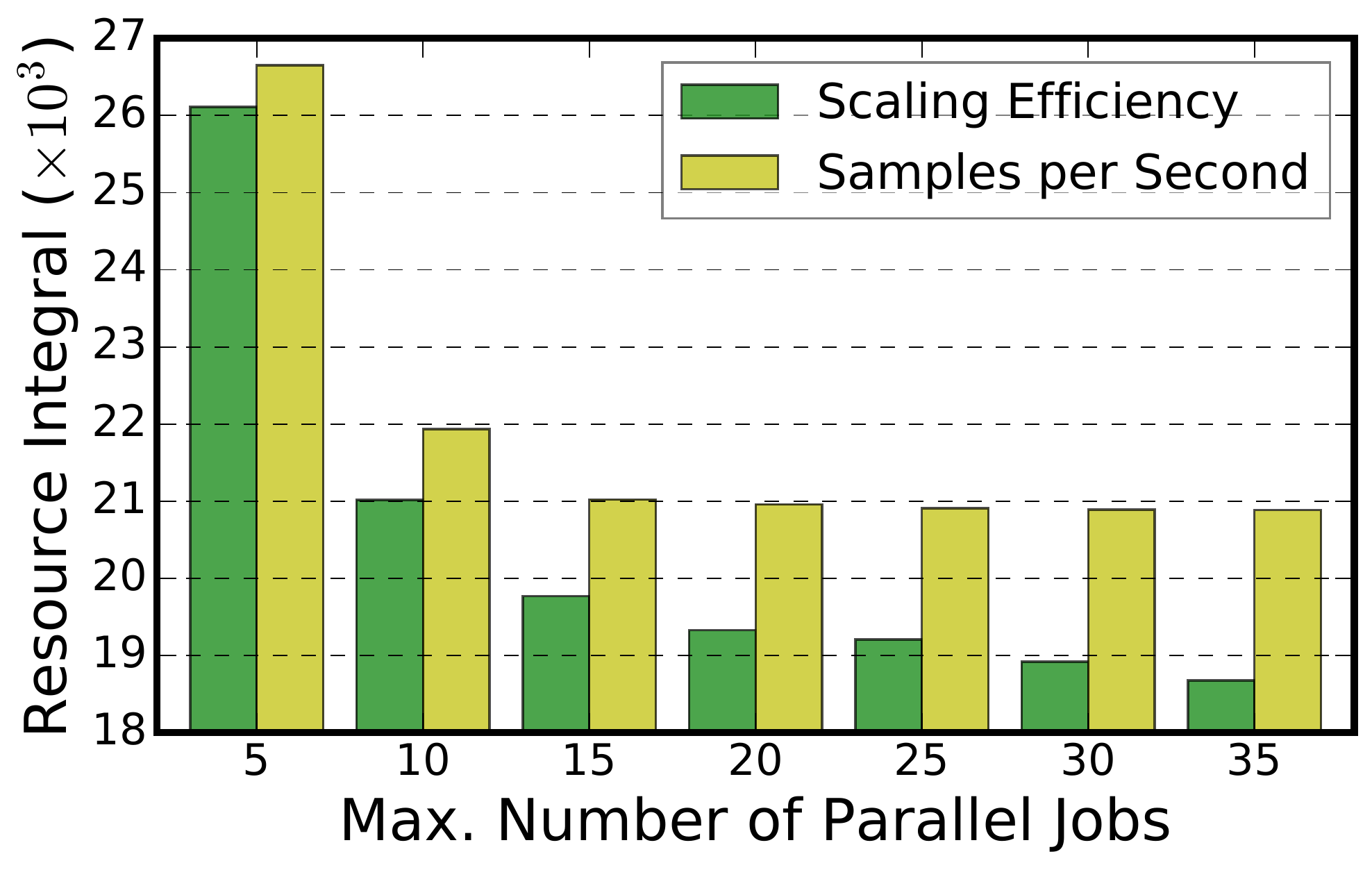}}}
\subfloat[]{\label{fig:pj-vs-runtime}{\includegraphics[width=0.33\linewidth]{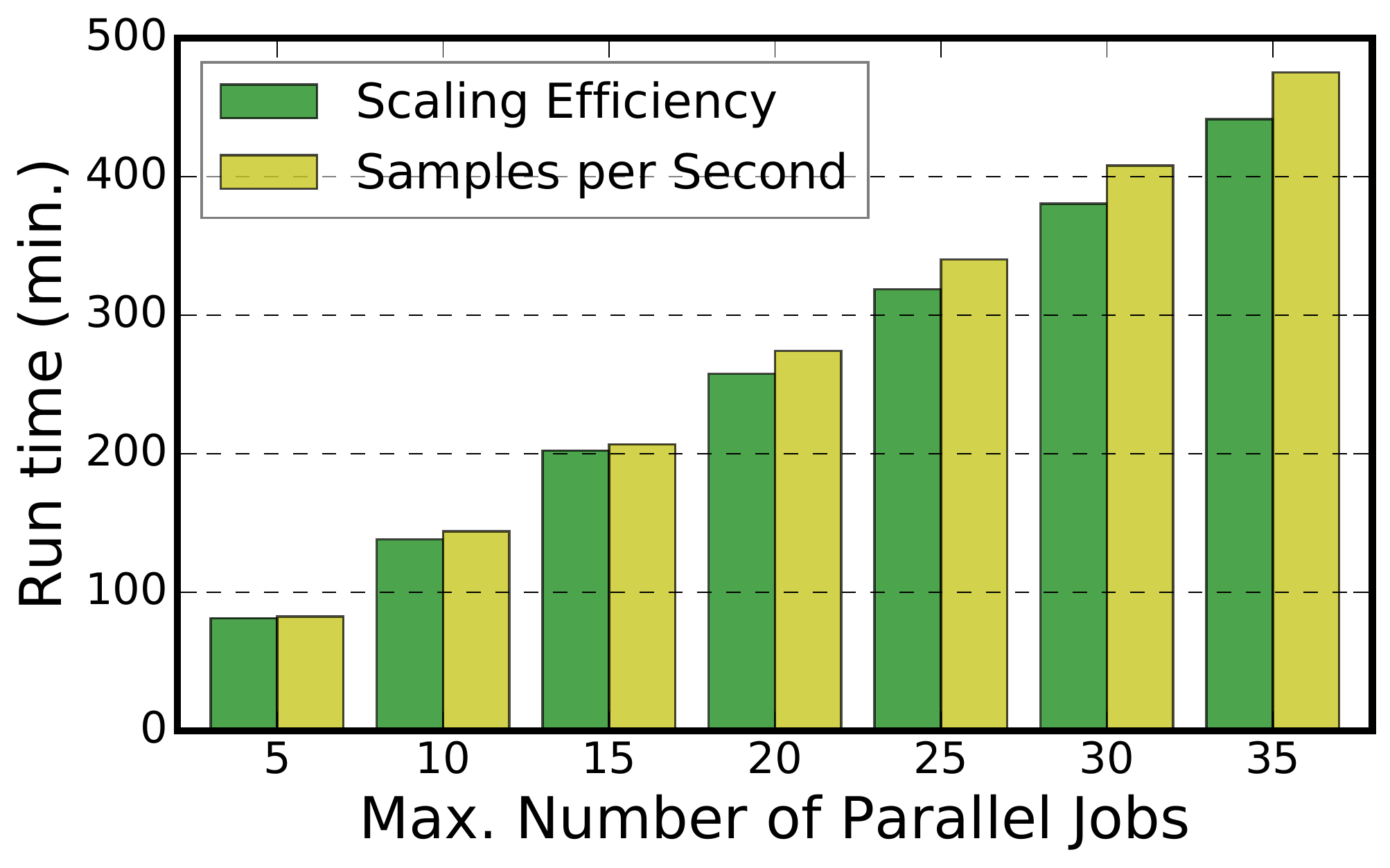}}}
\subfloat[]{\label{fig:pj-vs-effi}{\includegraphics[width=0.33\linewidth]{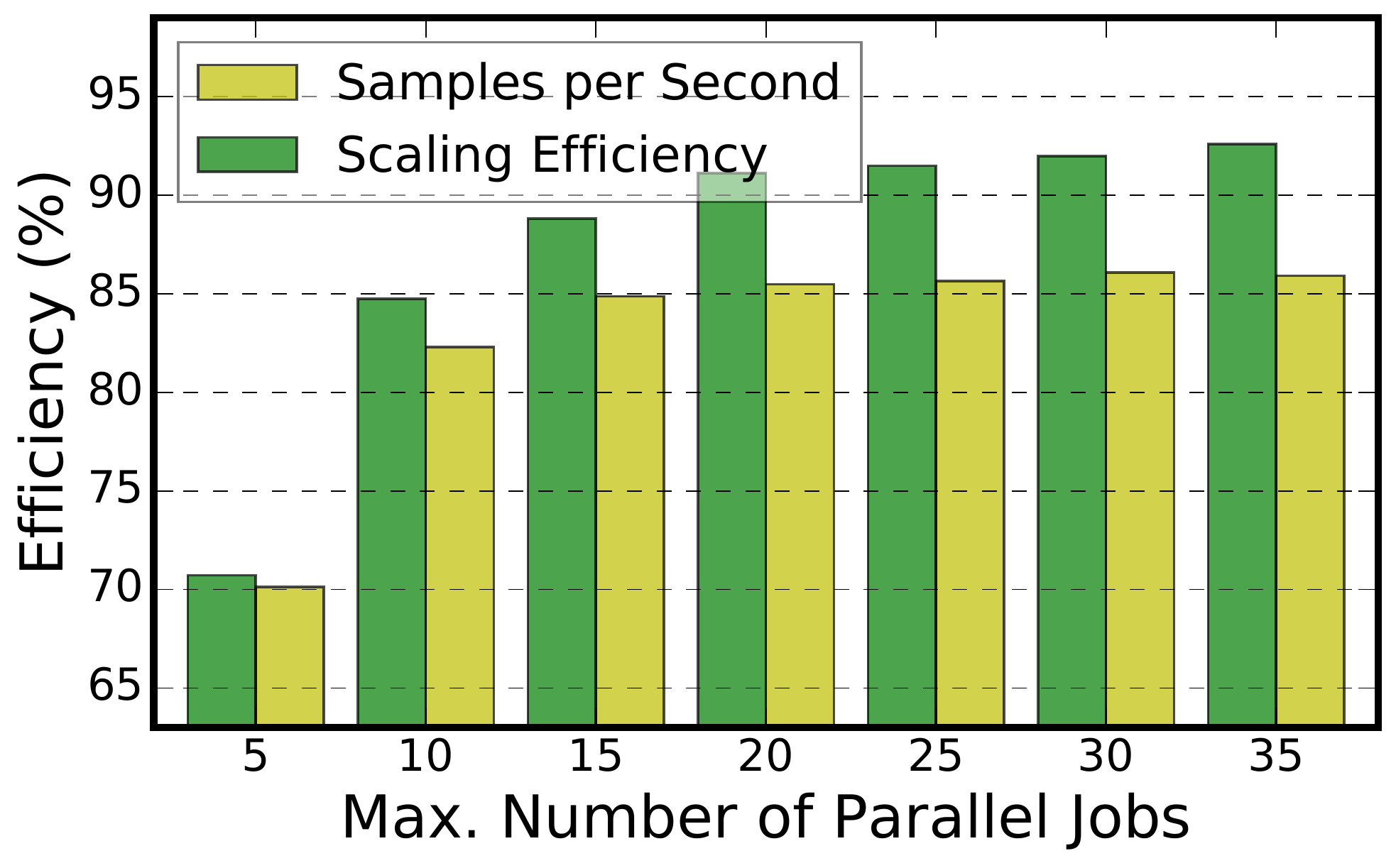}}}
\caption{Evaluation of the maximum parallel \bfjob{}s. We run 1000 \bfjob{}s composed with diverse DNNs, using different settings of the maximum parallel \bfjob{}s. Left: Comparison of resource consumption in \mt{}. Center: Average runtime of \bfjob{}. Right: Influence on resource utilization efficiency.} 
\label{fig:eva-maxJ}
\vspace{-.4cm}
\end{figure}

In conventional HPC scheduling, each job has a fixed size and thus the number of parallel running jobs, \pjmax{}, is determined by job sizes and computer size.
However, in \proj{} each job is elastic, and the number of nodes changes dynamically. 
The maximum number of parallel \bfjob{}s should in principle depend on the size of \pool{} and the minimum scale of the \bfjob{}s in the queue. 
If \bfjob{}s in the queue are treated as FCFS, there is no need for MILP to consider a number of \bfjob{}s more than the maximum, since doing so will cause serious job starvation~\cite{fan2021deep}. 
In this experiment we ran the two cases in \S\ref{sec:diverse-job} (i.e., diverse \bfjob{}s with two different objective metrics) with different \pjmax{}.
\autoref{fig:eva-maxJ} shows resource integrals, run time and average runtimes when different \pjmax{} are used.

As expected, if at a given time \pjmax{} increases, each \bfjob{}'size decreases because \pool{} is fixed.
Since \bfjob{}'s scaling efficiency decreases with \bfjob{} size, the resource utilization increases (\autoref{fig:pj-vs-effi}).
The resource integral (in aggregated \nodehour{} shown in \autoref{fig:pj-vs-mt}) decreases, because each application runs at a more efficient scale with a larger \pjmax{}.
However, as shown in \autoref{fig:pj-vs-runtime}, \bfjob{}'s runtime also increases significantly. 
The trade-off between the resource integral (or resource utilization efficiency) and \bfjob{} runtime is clear: for example, a 28\% decrease in the resource integral leads to a 442\% increase in average \bfjob{} runtime in this particular scenario when we increase the maximum number of parallel \bfjob{}s from 5 to 35.
In this experiment 12.4\% and 2.1\% of the machine time in the cases of 5 and 10 parallel \bfjob{}s, respectively, are unoccupied because the size of \pool{} is larger than the sum of the maximum sizes of all \bfjob{}s at a few events.
All other cases are fully occupied. 
This is why the resource integral for the \pjmax{}=5 case is significantly larger than others and the efficiency is much smaller than others.
Since the total computation demand are the same in all cases, the resource utilization efficiency increases with \pjmax{}, as shown in \autoref{fig:pj-vs-effi}.
This is because a larger \pjmax{} gives the scheduler opportunity to make more \bfjob{}s run at a more efficient size (i.e., smaller scale).

A proper number of parallel \bfjob{}s depends on the trade-off between the resource integral (directly related to resource utilization efficiency) 
and runtime of each individual \bfjob{}. 
For applications such as HPO using grid search, since the  runtime of each individual \bfjob{} does not make a difference but the resource integral makes a bigger difference, it is desirable to use a large number of parallel \bfjob{}s (e.g., upper bounded by maximum estimated \pool{} size and the minimum scale of each \bfjob{}).
However, in cases such as reinforcement learning or differential-method-based neural network architecture search~\cite{liu2018darts,vahdat2020unas,pham2018efficient} or heuristic algorithm-based HPO~\cite{jomaa2019hyp} where the next batch of \bfjob{}s depends on the outcome of the current \bfjob{}, a shorter \bfjob{} runtime is preferred and can be achieved by using less parallelism.

\begin{table}[htb]
\vspace{-3mm}
\centering
\caption{Average runtime (hours) of different \bfjob{}s with samples per second  used as the objective metric.}
\begin{tabular}{l|c|r|r|r|r|r|r}
\noalign{\hrule height 2pt}
\backslashbox{\bf DNN}{\bf \pjmax{}} & \multicolumn{1}{c|}{5} & \multicolumn{1}{c|}{10} & \multicolumn{1}{c|}{15} & \multicolumn{1}{c|}{20} & \multicolumn{1}{c|}{25} & \multicolumn{1}{c|}{30} & \multicolumn{1}{c}{35} \\\hline\hline
AlexNet    & 0.5 & 0.6 &  0.5 &  0.6 &  0.5 &  0.5 &  0.6 \\
ResNet18   & 0.4 & 0.4 &  0.4 &  0.4 &  0.5 &  0.5 &  0.5 \\
MnasNet    & 0.7 & 0.7 &  0.8 &  0.8 &  0.9 &  0.9 &  1.1 \\
MobileNets & 0.8 & 0.9 &  0.8 &  0.9 &  1.0 &  1.2 &  1.4 \\
ShuffleNet & 0.9 & 1.0 &  0.9 &  1.2 &  1.4 &  1.6 &  1.8 \\
VGG-16     & 2.3 & 3.4 &  4.1 &  5.4 &  6.1 &  6.3 &  7.4 \\
DenseNet   & 4.1 & 9.5 & 16.1 & 22.4 & 29.2 & 36.3 & 42.3 \\
\noalign{\hrule height 2pt}
\end{tabular}
\label{tbl:avg-rt-imgps}
\vspace{-3mm}
\end{table}

\autoref{tbl:avg-rt-imgps} compares average DNN runtimes for different \pjmax{} with samples per second as the objective metric.
As shown in \autoref{tbl:apps-benchmark}, AlexNet and DenseNet have the highest and lowest throughputs, respectively.
For easy comparison, \autoref{tbl:avg-rt-imgps} is ordered by DNN throughput.
Comparing DNNs by row, we see a clear increasing trend because MILP biased resources to high-throughput DNNs on the top.
Comparing by columns, we see that high-throughput DNNs can maintain the runtime regardless of \pjmax{} because they always get high priority, but the runtime of DenseNet (always lowest priority) increases significantly with \pjmax{}.

\begin{table}[htb]
\centering
\caption{Average runtime (in hours) of different \bfjob{} when scaling efficiency is used as objective metric.}
\begin{tabular}{l|c|r|r|r|r|r|r}
\noalign{\hrule height 2pt}
\backslashbox{\bf DNN}{\bf \pjmax{}} & \multicolumn{1}{c|}{5} & \multicolumn{1}{c|}{10} & \multicolumn{1}{c|}{15} & \multicolumn{1}{c|}{20} & \multicolumn{1}{c|}{25} & \multicolumn{1}{c|}{30} & \multicolumn{1}{c}{35} \\\hline\hline
VGG-16     & 1.98 & 2.8 & 3.3 & 3.8 & 4.5 & 5.0 &  5.2 \\
DenseNet   & 2.58 & 3.4 & 4.6 & 5.2 & 6.4 & 6.5 &  7.8 \\
ResNet18   & 0.77 & 1.1 & 1.5 & 1.9 & 2.1 & 2.4 &  2.6 \\
MobileNets & 0.89 & 1.8 & 2.7 & 3.8 & 4.7 & 5.9 &  7.4 \\
ShuffleNet & 1.15 & 2.1 & 2.9 & 4.2 & 5.4 & 7.2 &  8.5 \\
MnasNet    & 1.15 & 2.3 & 3.8 & 5.5 & 7.0 & 9.4 & 10.7 \\
AlexNet    & 0.94 & 2.4 & 3.9 & 5.3 & 6.8 & 7.6 &  8.8 \\
\noalign{\hrule height 2pt}
\end{tabular}
\label{tbl:avg-rt-spdup}
\vspace{-3mm}
\end{table}

Similarly, \autoref{tbl:avg-rt-spdup} compares the average runtime of different DNNs for different \pjmax{} with scaling efficiency as the objective metric.
Comparing with \autoref{tbl:avg-rt-imgps}, we see that different DNNs get relatively similar runtimes for each \pjmax{}.
AlexNet has the worst scaling efficiency and VGG-16 is the best according to \autoref{tbl:apps-benchmark}.
Lower priorities are given to AlexNet, especially for large \pjmax{} where fewer resources are allocated to each \bfjob{}, and thus \bfjob{}s with bad scalability are starved more.
DNNs with bad scalability are more sensitive to \pjmax{}. For example, the runtime of AlexNet increases nearly 10$\times$ (i.e., starved) while VGG-16 increases only 2.6$\times$ when \pjmax{} increases from 5 to 35.
Therefore, using a large \pjmax{} can increase the resource utilization and thus reduce the resource integral (\autoref{fig:pj-vs-mt}), but it also starves \bfjob{}.

\begin{obsbox}[boxsep=1pt,left=4pt,right=0pt,top=0pt,bottom=0pt]{green}{green}
\obs{Because \bfjob{}s are malleable, the number of jobs that can be run in parallel on a given number of nodes is flexible.
Greater parallelism enables more efficient resource use but lengthens the runtimes, especially for \bfjob{}s that get low priority based on the objective metric. 
\pjmax{} needs to be adjusted depending on the particular goal(s) of a resource provider or user.
}
\end{obsbox}

\subsection{Further discussion of diverse \bfjobs{}}
Two things of \bfjobs{} determine if they can be efficiently run using \proj{}: scalability and rescaling cost.
In this section we design experiments to evaluate the efficiency of \proj{} by varying the scalability and rescaling cost of \bfjobs{} respectively. 

\subsubsection{Scalability}\label{sec:scalability}
The scalability of a DNN depends, according to Amdahl's law~\cite{Amdahl67}, on the relative time spent in computation and communication, assuming that file I/O is not the bottleneck.
For example, different models for the same ImageNet classification problem have quite different scalabilities, as shown in \autoref{tbl:apps-benchmark}.

\begin{figure}[htb]
\vspace{-.3cm}
\centering
\includegraphics[width=0.8\columnwidth]{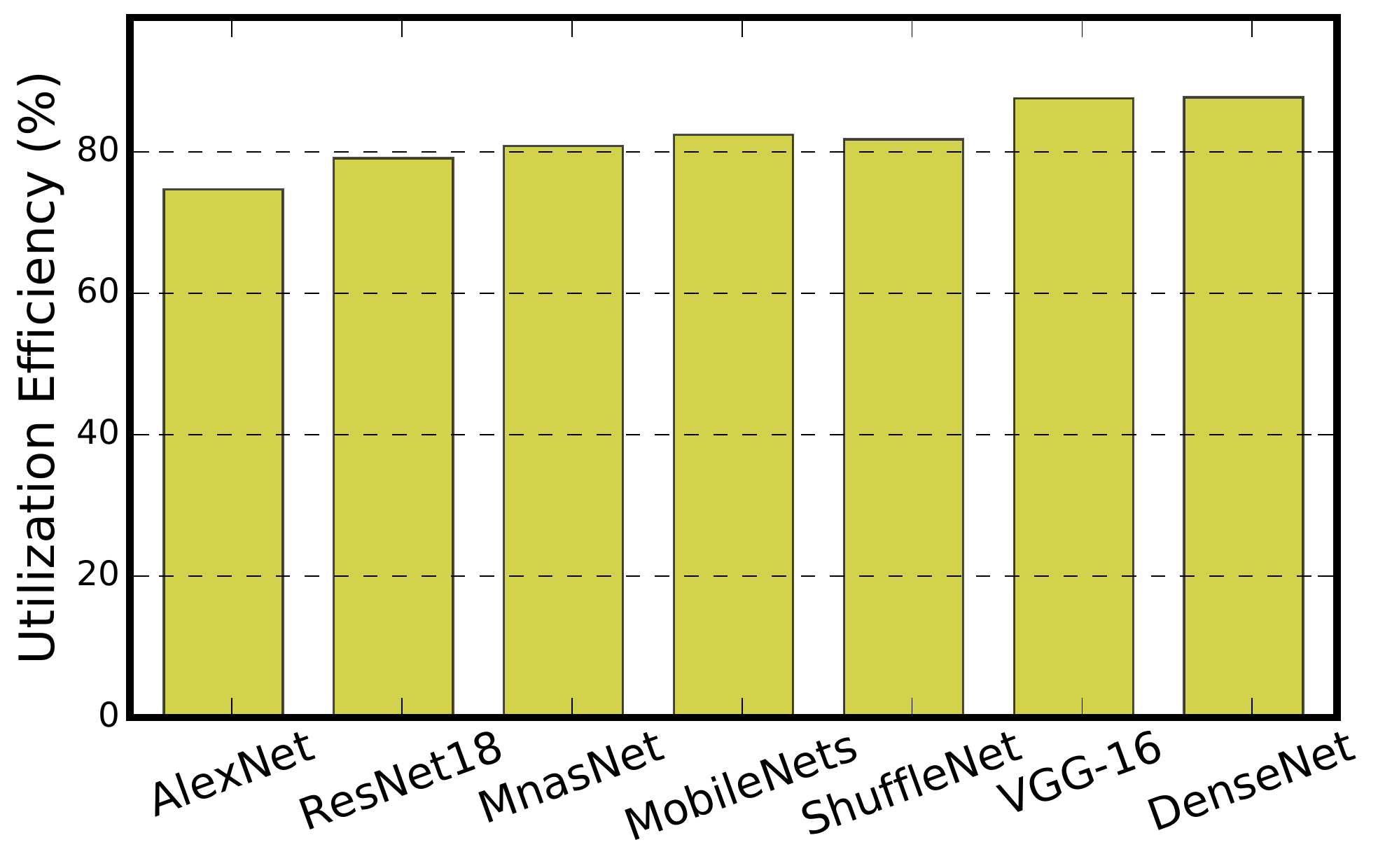}
\caption{A comparison of resource utilization efficiency when perform HPO for models with different scalability. DNN scaling efficiency increases from left to right.} 
\label{fig:var-hpt-eff-cmp}
\vspace{-.4cm}
\end{figure}

In order to evaluate \proj{}'s efficiency $U$ for DNNs with different scalabilities, we run the HPO experiment for each DNN shown in \autoref{tbl:apps-benchmark}.
As the shortest-running (AlexNet) only took about 70 hours to complete, we compute overall resource utilizations for each DNN over just the first 60 hours, so that all see the same resource availability. Our results are in \autoref{fig:var-hpt-eff-cmp}.

We see that all DNNs studied achieve $>$75\% resource utilization with BFTrainer,
with utilization efficiency increasing slightly with DNN scalability, from 75\% for AlexNet to 83\% for DenseNet. 
These results imply that optimizations that improve the scalability of individual \bfjobs{} are also beneficial to \proj{}. We note that while DNN training tasks that are less scalable on dedicated nodes are also less efficient with \proj{},
due to their less efficient use of larger resources,
we see no evidence that they perform relatively worse than more scalable DNNs when run with \proj{} rather than on dedicated nodes. 


\subsubsection{Rescaling cost}
\begin{figure}[htb]
\vspace{-.3cm}
\centering
\includegraphics[width=0.8\columnwidth]{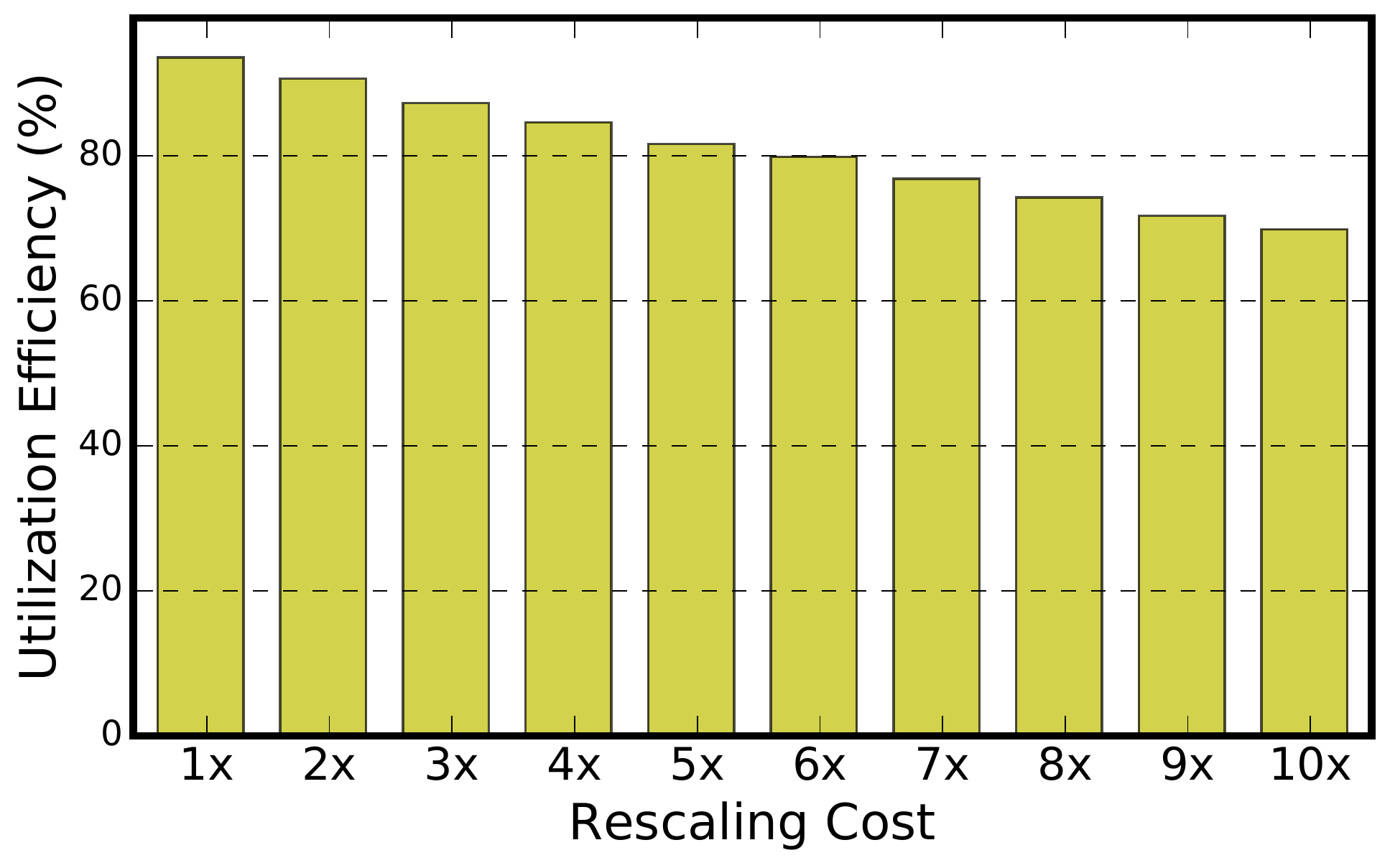}
\caption{A comparison of resource utilization efficiency with different artificial rescaling costs.} 
\label{fig:rex-hpt-eff-cmp}
\vspace{-.4cm}
\end{figure}

\proj{} efficiency is also influenced by \bfjob{}'s rescaling costs.
For example, as illustrated in \autoref{fig:decision}, scaling up from $n_j=4$ to $n_j=7$ involves an investment of $R_{up}$, which becomes profitable only after time $T_p$.
A larger $R_{up}$ leads to a later $T_p$, and as a \bfjob{} can be preempted at any time, further increases the chance of preemption (and thus a negative return).
If $R_{up}$ is large, the MILP will scale up only if it can do so by a large factor; 
if not, it will keep available nodes idle, leading to a lower $U$.
To study the influence of rescaling cost on utilization efficiency, we repeat the experiment of \S\ref{sec:hpt-exp-res} while artificially increase rescaling costs by from 2 to 10$\times$.
We present our results in \autoref{fig:rex-hpt-eff-cmp}.
As one can see, the resource utilization efficiency achieved by \proj{} decreased slightly with increased rescaling cost but much sublinearly. 

\begin{obsbox}[boxsep=1pt,left=4pt,right=0pt,top=0pt,bottom=0pt]{green}{green}
\obs{\proj{} is agnostic to DNN tasks. 
The scalability and rescaling cost of a particular DNN determine if it can be run efficiently using \proj{}.
DNNs with bad scalability is not appropriate for distributed training neither for \proj{}.
DNNs with more costly rescaling, including cost of cache missing in data pipeline because of rescaling, will lower the resource utilization efficiency of \proj{}.
}
\end{obsbox}

\section{Related Work}\label{sec:related-work}
Researchers have developed methods for running preemptable jobs on otherwise idle computers~\cite{anderson2002seti,thain2005distributed,marshall2011improving};
others have investigated methods for backfilling non-preemptable jobs in batch schedulers~\cite{zotkin1999job,talby1999supporting,srinivasan2002characterization,zhang2003integrated}, but always in the context of arbitrary, typically nonmalleable jobs.
\proj{}, in contrast, leverages specialized properties of DNN training jobs to make efficient use of otherwise unusable idle resources. 

Rodrigo et al.~\cite{rodrigo2015a2l2} proposed an HPC application-aware scheduling model to enable flexible backfilling of data-intensive applications by making use of dynamically malleable applications. 
Their survey of current scheduling challenges showed that malleable jobs are becoming increasingly common but that few HPC schedulers can exploit their particular characteristics.

The AdaptDL~\cite{qiao2020pollux} resource-adaptive DNN training and scheduling framework aims to make distributed deep learning easy and efficient in dynamic-resource environments such as shared clusters and the cloud. 
It uses Kubernetes~\cite{kubernetes} for scheduling, rescaling, and reconfiguring job batch size and learning rate to optimize training performance and resource utilization.
In contrast to \proj{}, it works actively but not preemptively.

Pilot-job systems (e.g., \cite{frey2002condor,raicu2007falkon,sfiligoi2009pilot,luckow2010saga,rubio2015gwpilot}, and this comprehensive survey~\cite{turilli2018comprehensive}), which submit placeholder jobs to queues, have long been used to enable flexible use of HPC resources. Some such systems rescale jobs to make efficient use of these computing nodes~\cite{berman1996application}. In contrast, \proj{} does not provision and hold  nodes; instead, it passively uses whatever nodes cannot be used by other jobs in the main batch queue.

\section{Conclusion and Future work}\label{sec:conclusion}
\proj{} makes optimal use of idle supercomputer nodes that cannot be, or are not, backfilled, to run applications (\bfjobs{}) that are (1) runnable at a range of scales and (2) re-scalable at modest cost.
It does so while optimizing a performance metric of choice by using mixed integer linear programming to reallocate resources for \bfjobs{} each time there is a change in idle resources or the number of \bfjobs{}. 
We focused in this proposal on one particular class  of \bfjobs{}, deep neural network training, due to their growing importance at many supercomputer centers.  
In future work, we will explore the application of \proj{} to other applications, and also investigate how to incorporate supercomputer network topology into the resource allocation algorithm, so as to consider the location of idle nodes when allocating to tasks.

\bibliographystyle{ACM-Reference-Format}
\bibliography{scavenger}

\section*{Government License}
The submitted manuscript has been created by UChicago Argonne, LLC, Operator of Argonne National Laboratory (``Argonne''). Argonne, a U.S.\ Department of Energy Office of Science laboratory, is operated under Contract No.\ DE-AC02-06CH11357. The U.S.\ Government retains for itself, and others acting on its behalf, a paid-up nonexclusive, irrevocable worldwide license in said article to reproduce, prepare derivative works, distribute copies to the public, and perform publicly and display publicly, by or on behalf of the Government.  The Department of Energy will provide public access to these results of federally sponsored research in accordance with the DOE Public Access Plan. http://energy.gov/downloads/doe-public-access-plan.

\end{document}